\documentstyle[aps,preprint,tighten,floats,psfig]{revtex}

\begin{document}
\draft
\preprint{\vbox{\it Submitted to Phys. Rev. D \hfill\rm CU-NPL-1152}}

\title{Determination of Decuplet Baryon Magnetic Moments \\
from QCD Sum Rules}
\author{Frank X. Lee}
\address{Nuclear Physics Laboratory, Department of Physics,
University of Colorado, \\Boulder, CO 80309-0446}
\date{\today}
\maketitle

\begin{abstract}
A complete set of QCD sum rules for the magnetic moments of 
decuplet baryons are derived using the external field method.
They are analyzed thoroughly using a Monte-Carlo based procedure.
Valid sum rules are identified 
under the criteria of OPE convergence and ground state dominance
and their predictions are obtained. 
The performances of these sum rules are further compared
and a favorable sum rule is designated for each member.
Correlations between the input and the output parameters are examined 
and large sensitivities to the quark condensate magnetic 
susceptibility $\chi$ are found.
Using realistic estimates of the QCD input parameters, 
the uncertainties on the magnetic moments are found relatively large 
and they can be attributed mostly to the poorly-known $\chi$.
It is shown that the accuracy can be improved to the 30\% level, 
provided the uncertainties in the QCD input parameters
can be determined to the 10\% level.
The computed magnetic moments are consistent with existing data. 
Comparisons with other calculations are made.

\end{abstract}
\vspace{1cm}
\parskip=2mm

\section{Introduction}
\label{intro}
The QCD sum rule method~\cite{SVZ79} has proven a powerful tool in
revealing the deep connection between hadron phenomenology and 
QCD vacuum structure via a few condensate parameters.
The method has been successfully applied to a variety
of problems to gain a field-theoretical understanding 
into the structure of hadrons.
Calculations of the nucleon magnetic moments 
in the approach were first carried out 
in Refs.~\cite{Ioffe84} and~\cite{Balitsky83}.
They were later refined and extended to the entire baryon octet in 
Refs.~\cite{Chiu86,Pasupathy86,Wilson87,Chiu87}.
On the other hand, the magnetic moments of decuplet baryons 
were less well studied within the same approach.
There were previous, unpublished reports in Ref.~\cite{Bely84} 
on $\Delta^{++}$ and $\Omega^-$ magnetic moments.
The magnetic form factor of $\Delta^{++}$
in the low $Q^2$ region was calculated based on a rather different 
technique~\cite{Bely93}.
In recent years, the magnetic moment of $\Omega^-$ has been measured
with remarkable accuracy~\cite{Wallace95}:
$\mu_{\scriptscriptstyle \Omega^-}=-2.02\pm0.05\;\mu_{\scriptscriptstyle N}$. 
The magnetic moment of $\Delta^{++}$ has also been extracted from 
pion bremsstrahlung~\cite{Bosshard91}:
$\mu_{\scriptscriptstyle \Delta^{++}}=4.5\pm 1.0\;\mu_{\scriptscriptstyle N}$.
In an earlier work~\cite{Heller87},
the magnetic moment of $\Delta^{0}$ extracted from 
$\pi^-p$ bremsstrahlung data was found to be consistent with
$\mu_{\scriptscriptstyle \Delta^{0}}=0$.
The experimental information provides new incentives for 
theoretical scrutiny of these observables.

In this work, we present a systematic, independent calculation of the 
magnetic moments for the entire decuplet family
in the QCD sum rule approach.
The goal is two-fold. 
First, we want to find out if the approach can be successfully 
applied to these observables by carrying out an explicit calculation.
Second, we want to achieve some realistic understanding of the 
uncertainties involved in such a determination by 
employing a Monte-Carlo based analysis procedure. 
This would help us assess the limitations and find ways for improvements.

We will show that both goals are achieved in this work.
The entire calculation is more challenging than the octet 
case due to the more complex spin structure of spin-3/2 particles.
One has to overcome enormous amount of algebra to arrive at the 
final results. 
But conceptually it presents no apparent difficulties. 
Particular attention is paid to the complete treatment of the 
phenomenological representation,
which leads to the isolation of the tensor structures from which 
the QCD sum rules for the magnetic moments can be constructed.
Flavor symmetry breakings in the strange quark are
treated consistently across the decuplet family.
The success also hinges upon a new analysis of the two-point 
functions~\cite{Lee97a}, which provides more accurately 
determined current couplings for normalization.
Part of the results on $\Delta^{++}$ and $\Omega^-$ 
have been communicated in a letter~\cite{Lee97b}.

Magnetic moments of decuplet baryons have also been studied in 
various other methods, including 
lattice QCD~\cite{Derek92},
chiral perturbation theory~\cite{Butler94},
Bethe-Salpeter formalism~\cite{Mitra84},
non-relativistic quark model~\cite{PDG92},
relativistic quark models~\cite{Schlumpf93,Capstick96,Linde95,Chao90,Georgi83},
chiral quark-soliton model~\cite{Kim97},
chiral bag model~\cite{Hong94},
cloudy bag model~\cite{Kriv87},
Skyrme model~\cite{Kim89}.
A comparison will be made with some of the calculations and with existing data.

Sec.~\ref{meth} deals with the derivation of the QCD sum rules.
Sec.~\ref{ana} discusses the Monte-Carlo analysis procedure.
Sec.~\ref{res} gives the results and discussions.
Sec.~\ref{sumcon} contains the conclusions.
The Appendix collects the QCD sum rules derived.

\section{Method}
\label{meth}

Consider the time-ordered two-point correlation function in the QCD vacuum
in the presence of a {\em constant} background 
electromagnetic field $F_{\mu\nu}$: 
\begin{equation}
\Pi_{\alpha\beta}(p)=i\int d^4x\; e^{ip\cdot x}
\langle 0\,|\, T\{\;\eta_{\alpha}(x)\,
\bar{\eta}_{\beta}(0)\;\}\,|\,0\rangle_F,
\label{cf2pt}
\end{equation}
where $\eta_{\alpha}$ is the interpolating field for the propagating baryon.
The subscript $F$ means that the correlation function 
is to be evaluated with an electromagnetic interaction term added 
to the QCD Lagrangian:
\begin{equation}
{\cal L}_I = - A_\mu J^\mu,
\end{equation}
where $A_\mu$ is the external electromagnetic potential and 
$J^\mu=e_q \bar{q} \gamma^\mu q$ the quark electromagnetic current. 

Since the external field can be made arbitrarily small, 
one can expand the correlation function
\begin{equation}
\Pi_{\alpha\beta}(p)=\Pi^{(0)}_{\alpha\beta}(p)
+\Pi^{(1)}_{\alpha\beta}(p)+\cdots.
\end{equation}
Here $\Pi^{(0)}_{\alpha\beta}(p)$ is the correlation function in 
the absence of the field, and gives rise to the mass sum rules of 
the baryons.
The magnetic moments will be extracted from the
QCD sum rules obtained from the linear response function 
$\Pi^{(1)}_{\alpha\beta}(p)$.  

The action of the external electromagnetic field 
is two-fold: it couples directly to the quarks in the baryon interpolating
fields,  and it also polarizes the QCD vacuum. 
The latter can be described by
introducing new parameters called  vacuum susceptibilities.

The interpolating field is constructed from quark fields, and 
has the quantum numbers of the baryon under consideration.
We use the following interpolating fields for the baryon decuplet family:
\begin{equation}
\begin{array}{l}
\eta_{\alpha}^{\Delta^{++}}=
\epsilon^{abc}\left(u^{aT}C\gamma_\alpha u^b\right) u^c,
\\
\eta_{\alpha}^{\Delta^+}=
\sqrt{1/3}\;\epsilon^{abc}
\left[2\left(u^{aT}C\gamma_\alpha d^b\right) u^c
+\left(u^{aT}C\gamma_\alpha u^b\right) d^c\right],
\\
\eta_{\alpha}^{\Delta^0}=
\sqrt{1/3}\;\epsilon^{abc}
\left[2\left(d^{aT}C\gamma_\alpha u^b\right) d^c
+\left(d^{aT}C\gamma_\alpha d^b\right) u^c\right],
\\
\eta_{\alpha}^{\Delta^-}=
\epsilon^{abc}\left(d^{aT}C\gamma_\alpha d^b\right) d^c,
\\
\eta_{\alpha}^{{\Sigma^*}^+}=
\sqrt{1/3}\;\epsilon^{abc}
\left[2\left(u^{aT}C\gamma_\alpha s^b\right) u^c
+\left(u^{aT}C\gamma_\alpha u^b\right) s^c\right],
\\
\eta_{\alpha}^{{\Sigma^*}^0}=
\sqrt{2/3}\;\epsilon^{abc}
\left[2\left(u^{aT}C\gamma_\alpha d^b\right) s^c
+\left(d^{aT}C\gamma_\alpha s^b\right) u^c
+\left(s^{aT}C\gamma_\alpha u^b\right) d^c\right],
\\
\eta_{\alpha}^{{\Sigma^*}^-}=
\sqrt{1/3}\;\epsilon^{abc}
\left[2\left(d^{aT}C\gamma_\alpha s^b\right) d^c
+\left(d^{aT}C\gamma_\alpha d^b\right) s^c\right],
\\
\eta_{\alpha}^{{\Xi^*}^0}=
\sqrt{1/3}\;\epsilon^{abc}
\left[2\left(s^{aT}C\gamma_\alpha u^b\right) s^c
+\left(s^{aT}C\gamma_\alpha s^b\right) u^c\right],
\\
\eta_{\alpha}^{{\Xi^*}^-}=
\sqrt{1/3}\;\epsilon^{abc}
\left[2\left(s^{aT}C\gamma_\alpha d^b\right) s^c
+\left(s^{aT}C\gamma_\alpha s^b\right) d^c\right],
\\
\eta_{\alpha}^{\Omega^-}=
\epsilon^{abc}\left(s^{aT}C\gamma_\alpha s^b\right) s^c.
\end{array}
\end{equation}
Here implicit function forms $\eta(x)$ and $q(x)$ (q=u,d,s) are assumed.
$C$ is the charge conjugation operator. The superscript $T$ means 
transpose. The indices a, b and c are color indices running from one to 
three. The antisymmetric tensor $\epsilon^{abc}$ ensures the three 
quarks form a color singlet state.
The normalization factors are chosen so that correlation functions 
of these interpolating fields coincide with each other under SU(3)-flavor 
symmetry (see Eqs.~(\ref{maso}) to (\ref{mass0})).

The interpolating field excites (or annihilates) the ground state 
as well as the excited states of the baryon 
from the QCD vacuum. The ability of a interpolating field to 
annihilate the {\em ground state} baryon into the QCD vacuum 
is described by a phenomenological parameter $\lambda_B$ (called current
coupling or pole residue), defined by the overlap
\begin{equation}
\langle 0\,|\,\eta_{\alpha}\,|\,Bps\rangle
=\lambda_B\,u_{\alpha}(p,s),
\end{equation}
where $u_{\alpha}$ is the Rarita-Schwinger spin-vector~~\cite{Rarita41}.

\subsection{Phenomenological Representation}
\label{rhs}

On the hadronic level, let us consider the linear response
defined by
\begin{equation}
\Pi^{(1)}_{\alpha\beta}(p) =
i\int d^4x\; e^{ip\cdot x}\langle 0\,|\,
\eta_{\alpha}(x)\,
\left[ -i \int d^4y\;A_\mu(y) J^\mu(y)\right]
\bar{\eta}_{\beta}(0)\,|\,0\rangle.
\label{phen1}
\end{equation}
After inserting two complete sets of physical intermediate states, it becomes
\begin{eqnarray}
\Pi^{(1)}_{\alpha\beta}(p) & = &
\int d^4x \int d^4y {d^4 k \over (2\pi)^4} {d^4 k^\prime \over (2\pi)^4}
\sum_{BB^\prime} \sum_{ss^\prime} 
{-i \over k^2-M^2_B -i\epsilon}\;
{-i \over {k^\prime}^2-M^2_{B^\prime} -i\epsilon}
\nonumber \\ & & 
e^{ip\cdot x} A_\mu(y) 
\langle 0 | \eta_{\alpha}(x) | ks \rangle 
\langle ks | J^\mu(y) | k^\prime s^\prime \rangle 
\langle k^\prime s^\prime | \bar{\eta}_{\beta}(0) | 0 \rangle.
\label{phen2}
\end{eqnarray}
QCD sum rule calculations are most conveniently done in the fixed-point gauge.
For electromagnetic field, it is defined by $x_\mu A^\mu(x)=0$.
In this gauge, the electromagnetic potential is given by 
\begin{equation}
A_\mu(y)=-{1\over 2} F_{\mu\nu} y^\nu.
\end{equation}
The electromagnetic vertex of spin-3/2 baryons 
is defined by the current matrix element~\cite{Derek92}
\begin{equation}
\langle ks | J^\mu(0) | k^\prime s^\prime \rangle = 
\bar{u}_\alpha(k,s)\;{\cal O}^{\alpha \mu \beta}(P,q)\; 
u_\beta(k^\prime,s^\prime).
\end{equation}
The Lorentz covariant tensor
\begin{equation}
{\cal O}^{\alpha \mu \beta}(P,q) \equiv
-g^{\alpha\beta} \left( a_1 \gamma^\mu + {a_2 \over 2M_B} P^\mu \right)
- {q^\alpha q^\beta \over (2M_B)^2} 
\left( c_1 \gamma^\mu + {c_2 \over 2M_B} P^\mu \right),
\end{equation}
where $P=k+k^\prime$ and $q=k-k^\prime$,
satisfies the standard requirements of invariance under time reversal,
parity, $G$ parity, and gauge transformations. The parameters $a_1$, 
$a_2$, $c_1$ and $c_2$ are independent covariant vertex functions.
They are related to the multipole form factors by
\begin{equation}
\begin{array}{l}
G_{E0}(q^2) = (1+{2\over 3} \tau) \left[ a_1 + (1+\tau) a_2 \right]
-{1\over 3} \tau (1+\tau) \left[ c_1 + (1+\tau) c_2 \right]
\\
G_{E2}(q^2) = \left[ a_1 + (1+\tau) a_2 \right]
-{1\over 2} (1+\tau) \left[ c_1 + (1+\tau) c_2 \right]
\\
G_{M1}(q^2) = (1+{4\over 5}\tau) a_1 - {2\over 5}\tau (1+\tau) c_1
\\
G_{M3}(q^2) = a_1 - {1\over 2} (1+\tau) c_1.
\end{array}
\label{multipole}
\end{equation}
where $\tau=-q^2/(2M_B)^2 (\ge 0)$.
They are referred to as charge (E0), electric quadrupole (E2), 
magnetic dipole (M1), and magnetic octupole (M3) form factors.
The magnetic moment is related to the magnetic dipole form 
factor $G_{M1}(q^2)$ at zero momentum transfer.
From Eq.~(\ref{multipole}), it is clear that 
\begin{equation}
G_{M1}(0)=a_1 \equiv \mu_{\scriptscriptstyle B},
\end{equation}
where the magnetic moment $\mu_{\scriptscriptstyle B}$ 
is in units of particle's natural magneton: $e\hbar/(2c M_B)$.
So the goal is to isolate terms in Eq.~(\ref{phen2}) 
that involve only $a_1$.

The ground state contribution to Eq.~(\ref{phen2}) can be written as 
\begin{eqnarray}
\Pi^{(1)}_{\alpha\beta}(p) & = & {i\over 2} \lambda^2_B F_{\mu\nu} 
\int d^4x {d^4 k \over (2\pi)^4} 
e^{i(p-k)\cdot x} {1 \over k^2-M^2_B -i\epsilon}\;
\sum_{s} u_\alpha(k,s) \bar{u}_\rho(k,s) 
\nonumber \\ & & 
{\partial \over \partial q^\nu} \left[
{1 \over (k-q)^2-M^2_B -i\epsilon}
O^{\rho\mu\lambda}(2k-q,q)
\sum_{s^\prime} u_\lambda(k-q,s^\prime) \bar{u}_\beta(k-q,s^\prime) 
\right] \bigg|_{q=0}.
\label{phen3}
\end{eqnarray}
In arriving at Eq.~(\ref{phen3}), we have used a number of steps:
the translation invariance on $\eta_\alpha(x)$ and $J^\mu(y)$,
a change of variable from $k^\prime$ to $q$,
the relation 
\begin{equation}
\int d^4y e^{iq\cdot y} y^\nu = -i (2\pi)^4 
{\partial \over \partial q^\nu} \delta^4(q),
\end{equation}
integration by parts,
and the Rarita-Swinger spin sum~\cite{Rarita41}
\begin{equation}
\sum_{s} u_\alpha (p,s) \bar{u}_\beta (p,s) =
-(\hat{p} + M_B) \left(
g_{\alpha\beta} - {1\over3} \gamma_\alpha \gamma_\beta 
- {2 p_\alpha p_\beta \over 3 M^2_B} 
+ {p_\alpha \gamma_\beta - p_\beta \gamma_\alpha \over 3 M_B} \right),
\end{equation}
with normalization $\bar{u}_\alpha u_\alpha = 2M_B$.
The hat notation denotes $\hat{p}=p^\alpha\,\gamma_\alpha$.

Direct evaluation of Eq.~(\ref{phen3}) leads to numerous tensor 
structures, not all of them are independent of each other.
The dependencies can be removed  by ordering the gamma matrices 
in a specific order. Here we choose to order 
in $ \hat{p} \gamma_\alpha \gamma_\mu \gamma_\nu \gamma_\beta $.
After a lengthy calculation, 18 tensor structures which involve only 
$a_1$ are isolated. They can be organized as 
\begin{equation}
\begin{array}{ll}
\Pi^{(1)}_{\alpha\beta}(p) =
\\
\;\;\; \mbox{WE}_1(p^2)\;\hat{p} F^{\mu\nu} \sigma_{\mu\nu} g_{\alpha\beta}
&
+ \mbox{WO}_1(p^2)\;F^{\mu\nu} \sigma_{\mu\nu} g_{\alpha\beta}
\\
+ \mbox{WE}_2(p^2)\;\hat{p} p_{\alpha} F^{\mu\nu} \sigma_{\mu\nu} p_{\beta}
&
+ \mbox{WO}_2(p^2)\;p_{\alpha} F^{\mu\nu} \sigma_{\mu\nu} p_{\beta}
\\
+ \mbox{WE}_3(p^2)\;\hat{p} \gamma_{\alpha} F^{\mu\nu} \sigma_{\mu\nu} \gamma_{\beta}
&
+ \mbox{WO}_3(p^2)\;\gamma_{\alpha} F^{\mu\nu} \sigma_{\mu\nu} \gamma_{\beta}
\\
+ \mbox{WE}_4(p^2)\; p_{\alpha} F^{\mu\nu} \sigma_{\mu\nu} \gamma_{\beta}
&
+ \mbox{WO}_4(p^2)\;\hat{p} p_{\alpha} F^{\mu\nu} \sigma_{\mu\nu} \gamma_{\beta}
\\
+ \mbox{WE}_5(p^2)\; \gamma_{\alpha} F^{\mu\nu} \sigma_{\mu\nu} p_{\beta}
&
+ \mbox{WO}_5(p^2)\;\hat{p} \gamma_{\alpha} F^{\mu\nu} \sigma_{\mu\nu} p_{\beta}
\\
+ \mbox{WE}_6(p^2)\;\hat{p} \gamma_\alpha F^{\mu\nu} 
	 (\gamma_\mu g_{\beta\nu} -\gamma_\nu g_{\beta\mu})
&
+ \mbox{WO}_6(p^2)\;\gamma_{\alpha}F^{\mu\nu} 
	 (\gamma_\mu g_{\beta\nu} -\gamma_\nu g_{\beta\mu})
\\
+ \mbox{WE}_7(p^2)\;\hat{p} F^{\mu\nu} 
	(\gamma_\mu g_{\alpha\nu} - \gamma_\nu g_{\alpha\mu}) \gamma_\beta
&
+ \mbox{WO}_7(p^2)\;F^{\mu\nu} 
	(\gamma_\mu g_{\alpha\nu} - \gamma_\nu g_{\alpha\mu}) \gamma_\beta
\\
+ \mbox{WE}_8(p^2)\; p_\alpha F^{\mu\nu} 
	(\gamma_\mu g_{\beta\nu} - \gamma_\nu g_{\beta\mu})
&
+ \mbox{WO}_8(p^2)\;\hat{p} p_\alpha F^{\mu\nu} 
	(\gamma_\mu g_{\beta\nu} - \gamma_\nu g_{\beta\mu})
\\
+ \mbox{WE}_9(p^2)\; F^{\mu\nu} 
	(\gamma_\mu g_{\alpha\nu} - \gamma_\nu g_{\alpha\mu}) p_\beta
&
+ \mbox{WO}_9(p^2)\;\hat{p} F^{\mu\nu} 
	(\gamma_\mu g_{\alpha\nu} - \gamma_\nu g_{\alpha\mu}) p_\beta
+ \cdots.
\end{array}
\label{phen4}
\end{equation}
The tensor structures associated with WE$_i$ have odd number of gamma matrices,
while those associated with WO$_i$ have even number of gamma matrices.
Apart from a common factor 
$i{\lambda}^2_B\;\mu_{\scriptscriptstyle B}/(p^2-M^2_B)^2$,
the invariant functions are given by
\begin{equation}
\begin{array}{ll}
\mbox{WE}_1={1\over 2}, & 
\mbox{WO}_1={1\over 2}M_B,\\ 
\mbox{WE}_2={-1\over 9M^2_B}, &
\mbox{WO}_2={-1\over 9M_B},\\
\mbox{WE}_3={-7\over 18},&
\mbox{WO}_3={-7\over 18} M_B,\\
\mbox{WE}_4={ 7\over 18},&
\mbox{WO}_4={ 7\over 18 M_B},\\
\mbox{WE}_5={-7\over 18}, &
\mbox{WO}_5={-7\over 18 M_B},\\
\mbox{WE}_6={ 2\over  3},&
\mbox{WO}_6={ 2\over  3} M_B,\\
\mbox{WE}_7={-2\over  3}, &
\mbox{WO}_7={-2\over  3} M_B,\\
\mbox{WE}_8={-2\over  3}, &
\mbox{WO}_8={-2\over  3 M_B},\\
\mbox{WE}_9={-2\over  3},&
\mbox{WO}_9={-2\over  3 M_B}.
\end{array}
\end{equation}

In addition to the ground state contribution, there exist also 
excited state contributions.
For a generic invariant function, the pole structure has the form
\begin{equation}
{\lambda^2_B\;\mu_{\scriptscriptstyle B} \over (p^2-M_B^2)^2}
+ \sum_{B^*}{C_{B\leftrightarrow B^*} \over (p^2-M_B^2)(p^2-M_{B^*}^2)}
+ \cdots.
\label{pole}
\end{equation}
where $C_{B\leftrightarrow B^*}$ are constants.
The first term is the ground state double pole which contains the 
desired magnetic moment of the baryon, the second term
represents the non-diagonal transitions between the ground state and the
excited states caused by the external field, and the ellipses
represent pure excited state contributions.  
Upon Borel transform, one has
\begin{equation}
{\lambda^2_B\;\mu_{\scriptscriptstyle B} \over M^2}\;e^{-M_B^2/M^2}
+ e^{-M_B^2/M^2} \left[\sum_{B^*}
{C_{B\rightarrow B^*} \over M_{B^*}^2 - M_B^2}
\left(1-e^{-(M_{B^*}^2-M_B^2)/M^2}\right)\right]
+ \cdots.
\label{pole-borel}
\end{equation}
We see that the transitions give rise to a contribution
that is not exponentially suppressed relative to the ground state.
This is a general feature of the external-field technique.
The strength of such transitions at each structure is
{\it a priori} unknown and is an additional source of contamination in
the determination of $\mu_B$ not found in mass sum rules.  The usual
treatment of the transitions is to approximate the quantity in the
square brackets by a constant, which is to
be extracted from the sum rule along with the ground state property of
interest. 
Inclusion of such contributions is necessary for the correct extraction 
of the magnetic moments.
The pure excited state contributions
are exponentially suppressed relative to the ground state and
can be modeled in the usual way by introducing a continuum model and
threshold parameter.

\subsection{Calculation of the QCD Side}
\label{lhs}

On the quark level, one evaluates the correlation function in 
Eq.~(\ref{cf2pt}) using Operator Product Expansion (OPE).
The calculation is most readily done in coordinate space.
To arrive at the final sum rules, 
one needs a subsequent Fourier transform, 
followed by a Borel transform.

We decide to carry out 
four separate calculations for $\Omega^-$(sss), ${\Sigma^*}^+$(uus),
${\Xi^*}^0$(uss), and ${\Sigma^*}^0$(uds).
They have distinct strange quark content, which requires special treatment.
The QCD sum rules for other members can be obtained by appropriate 
substitutions in those for these four members.
 
The master formula, which is obtained from contracting out the quark pairs
in the correlation function,  is given by, for $\Omega^-$:
\begin{equation}
\begin{array}{rll}
\langle 0\,|\, T\{\;\eta^{\Omega^-}_{\alpha}(x)\,
\bar{\eta}^{\Omega^-}_{\beta}(0)\;\}\,|\,0\rangle_F = 
\\ 
2 \epsilon^{abc}\epsilon^{a^\prime b^\prime c^\prime}
\{ &  
S^{aa^\prime}_s \mbox{Tr} \left[ \gamma_\beta C {S^{bb^\prime}_s}^T 
C \gamma_\alpha S^{cc^\prime}_s \right]
& 
+ 2 S^{aa^\prime}_s  \gamma_\beta C {S^{bb^\prime}_s}^T 
C \gamma_\alpha S^{cc^\prime}_s
\}, 
\end{array}
\label{maso}
\end{equation}
for ${\Sigma^*}^+$:
\begin{equation}
\begin{array}{lll}
\langle 0\,|\, T\{\;\eta^{{\Sigma^*}^+}_{\alpha}(x)\,
\bar{\eta}^{{\Sigma^*}^+}_{\beta}(0)\;\}\,|\,0\rangle_F = 
&  
{2 \over 3} \epsilon^{abc}\epsilon^{a^\prime b^\prime c^\prime}
\{ & \\  
\;\;\; S^{aa^\prime}_u \mbox{Tr} \left[ \gamma_\beta C {S^{bb^\prime}_u}^T 
C \gamma_\alpha S^{cc^\prime}_s \right]
&  
+ S^{aa^\prime}_u \mbox{Tr} \left[ \gamma_\beta C {S^{bb^\prime}_s}^T 
C \gamma_\alpha S^{cc^\prime}_u \right]
&
+ S^{aa^\prime}_s \mbox{Tr} \left[ \gamma_\beta C {S^{bb^\prime}_u}^T 
C \gamma_\alpha S^{cc^\prime}_u \right]
\\
+ 2 S^{aa^\prime}_u  \gamma_\beta C {S^{bb^\prime}_u}^T 
C \gamma_\alpha S^{cc^\prime}_s 
&
+ 2 S^{aa^\prime}_u  \gamma_\beta C {S^{bb^\prime}_s}^T 
C \gamma_\alpha S^{cc^\prime}_u
&
+ 2 S^{aa^\prime}_s \gamma_\beta C {S^{bb^\prime}_u}^T 
C \gamma_\alpha S^{cc^\prime}_u 
\},
\end{array}
\label{massp}
\end{equation}
for ${\Xi^*}^0$:
\begin{equation}
\begin{array}{lll}
\langle 0\,|\, T\{\;\eta^{{\Xi^*}^0}_{\alpha}(x)\,
\bar{\eta}^{{\Xi^*}^0}_{\beta}(0)\;\}\,|\,0\rangle_F = 
&  
{2 \over 3} \epsilon^{abc}\epsilon^{a^\prime b^\prime c^\prime}
\{ & \\  
\;\;\; S^{aa^\prime}_s \mbox{Tr} \left[ \gamma_\beta C {S^{bb^\prime}_s}^T 
C \gamma_\alpha S^{cc^\prime}_u \right]
&  
+ S^{aa^\prime}_s \mbox{Tr} \left[ \gamma_\beta C {S^{bb^\prime}_u}^T 
C \gamma_\alpha S^{cc^\prime}_s \right]
&
+ S^{aa^\prime}_u \mbox{Tr} \left[ \gamma_\beta C {S^{bb^\prime}_s}^T 
C \gamma_\alpha S^{cc^\prime}_s \right]
\\
+ 2 S^{aa^\prime}_s  \gamma_\beta C {S^{bb^\prime}_s}^T 
C \gamma_\alpha S^{cc^\prime}_u 
&
+ 2 S^{aa^\prime}_s  \gamma_\beta C {S^{bb^\prime}_u}^T 
C \gamma_\alpha S^{cc^\prime}_s
&
+ 2 S^{aa^\prime}_u \gamma_\beta C {S^{bb^\prime}_s}^T 
C \gamma_\alpha S^{cc^\prime}_s 
\},
\end{array}
\label{masx0}
\end{equation}
and for ${\Sigma^*}^0$:
\begin{equation}
\begin{array}{lll}
\langle 0\,|\, T\{\;\eta^{{\Sigma^*}^0}_{\alpha}(x)\,
\bar{\eta}^{{\Sigma^*}^0}_{\beta}(0)\;\}\,|\,0\rangle_F = 
&  
{2 \over 3} \epsilon^{abc}\epsilon^{a^\prime b^\prime c^\prime}
\{ & \\  
\;\;\; S^{aa^\prime}_u \mbox{Tr} \left[ \gamma_\beta C {S^{bb^\prime}_d}^T 
C \gamma_\alpha S^{cc^\prime}_s \right]
&  
+ S^{aa^\prime}_d \mbox{Tr} \left[ \gamma_\beta C {S^{bb^\prime}_s}^T 
C \gamma_\alpha S^{cc^\prime}_u \right]
&
+ S^{aa^\prime}_s \mbox{Tr} \left[ \gamma_\beta C {S^{bb^\prime}_u}^T 
C \gamma_\alpha S^{cc^\prime}_d \right]
\\
+ S^{aa^\prime}_u  \gamma_\beta C {S^{bb^\prime}_d}^T 
C \gamma_\alpha S^{cc^\prime}_s 
&
+ S^{aa^\prime}_d  \gamma_\beta C {S^{bb^\prime}_s}^T 
C \gamma_\alpha S^{cc^\prime}_u
&
+ S^{aa^\prime}_s \gamma_\beta C {S^{bb^\prime}_u}^T 
C \gamma_\alpha S^{cc^\prime}_d 
\\ 
+ S^{aa^\prime}_u  \gamma_\beta C {S^{bb^\prime}_s}^T 
C \gamma_\alpha S^{cc^\prime}_d 
&
+ S^{aa^\prime}_d  \gamma_\beta C {S^{bb^\prime}_u}^T 
C \gamma_\alpha S^{cc^\prime}_s
& 
+ S^{aa^\prime}_s \gamma_\beta C {S^{bb^\prime}_d}^T 
C \gamma_\alpha S^{cc^\prime}_u 
\}.
\end{array}
\label{mass0}
\end{equation}
In the above equations, 
\begin{equation}
S^{ab}_q (x,0;F) \equiv 
\langle 0\,|\, T\{\;q^a(x)\,
\bar{q}^b(0)\;\}\,|\,0\rangle_F,
 \hspace{3mm} q=u, d, s,
\end{equation}
is the fully interacting 
quark propagator in the presence of the electromagnetic field. 
To first order in $F_{\mu\nu}$ and $m_q$ 
(assume $m_u=m_d=0, m_s\neq 0$), and order $x^4$, it is given 
by~\cite{Ioffe84,Pasupathy86,Wilson87}:
\begin{eqnarray}
S^{ab}_q(x,0;Z) &\equiv& 
 {i \over 2\pi^2} {\hat{x}\over x^4} \delta^{ab}
- {m_q \over 4\pi^2 x^2} \delta^{ab}
- {1\over 12}\langle\bar{q}q\rangle \delta^{ab}
+ {im_q \over 48} \langle\bar{q}q\rangle \hat{x} \delta^{ab}
\nonumber \\ & &
+ {1\over 192} \langle\bar{q}g_c\sigma\cdot Gq\rangle x^2 \delta^{ab}
- {im_q\over 1152} \langle\bar{q}g_c\sigma\cdot Gq\rangle
	\hat{x} x^2 \delta^{ab}
- {1\over 3^3 2^{10}} \langle\bar{q}q\rangle \langle g^2_c G^2\rangle
	 x^4 \delta^{ab}
\nonumber \\ & &
+ {i\over 32\pi^2} (g_cG^n_{\alpha\beta}) 
  { \hat{x} \sigma^{\alpha\beta} +\sigma^{\alpha\beta} \hat{x}\over x^2 }
  \left({\lambda^n\over 2}\right)^{ab}
+ {1\over 48} {i\over 32\pi^2} \langle g^2_c G^2\rangle
  {\hat{x} \sigma^{\alpha\beta} +\sigma^{\alpha\beta} \hat{x}\over x^2} 
  \left({\lambda^n\over 2}\right)^{ab}
\nonumber \\ & &
+ {1\over 3^2 2^{10}} \langle\bar{q}q\rangle \langle g^2_c G^2\rangle
  x^2 \sigma^{\alpha\beta} \left({\lambda^n\over 2}\right)^{ab}
-  {1\over 192}\langle\bar{q}g_c\sigma\cdot Gq\rangle
  \sigma^{\alpha\beta} \left({\lambda^n\over 2}\right)^{ab}
\nonumber \\ & &
+  {im_q\over 768}\langle\bar{q}g_c\sigma\cdot Gq\rangle
  \left( \hat{x} \sigma^{\alpha\beta} +\sigma^{\alpha\beta} \hat{x} \right)
  \left({\lambda^n\over 2}\right)^{ab}
+  {i e_q\over 32\pi^2} F_{\alpha\beta}
  { \hat{x} \sigma^{\alpha\beta} +\sigma^{\alpha\beta} \hat{x} \over x^2 } 
  \delta^{ab}
\nonumber \\ & &
-  {e_q\over 24} \chi \langle\bar{q}q\rangle 
   F_{\alpha\beta} \sigma^{\alpha\beta} \delta^{ab}
+  {ie_q m_q\over 96} \chi \langle\bar{q}q\rangle F_{\alpha\beta}
  \left( \hat{x} \sigma^{\alpha\beta} +\sigma^{\alpha\beta} \hat{x} \right)
   \delta^{ab}
\nonumber \\ & &
+  {e_q \over 288} \langle\bar{q}q\rangle F_{\alpha\beta}
  \left( x^2 \sigma^{\alpha\beta} - 2 x_\rho x^\beta \sigma^{\beta\alpha} 
  \right) \delta^{ab}
\nonumber \\ & &
+  {e_q \over 576} \langle\bar{q}q\rangle F_{\alpha\beta}
  \left[ x^2 (\kappa+\xi) \sigma^{\alpha\beta} 
  - x_\rho x^\beta (2\kappa-\xi) \sigma^{\beta\alpha} \right] \delta^{ab}
\nonumber \\ & &
-  {e_q \over 16} \langle\bar{q}q\rangle 
   \left( \kappa F_{\alpha\beta} 
   - {i\over 4} \xi \epsilon_{\alpha\beta\mu\nu} F^{\mu\nu} \right)
  \left({\lambda^n\over 2}\right)^{ab}
+ \mbox{higher order terms}.
\label{prop}
\end{eqnarray}
We use the convention $\epsilon^{0123}=+1$ in this work.
The vacuum susceptibilities are defined by
\begin{equation}
\begin{array}{r}
\langle\bar{q} \sigma_{\mu\nu} q\rangle_F \equiv
e_q \chi \langle\bar{q}q\rangle F_{\mu\nu}, \\
\langle\bar{q} g_c G_{\mu\nu} q\rangle_F \equiv
e_q \kappa \langle\bar{q}q\rangle F_{\mu\nu}, \\
\langle\bar{q} g_c \epsilon_{\mu\nu\rho\lambda} G^{\rho\lambda} \gamma_5 
q\rangle_F \equiv
i e_q \xi \langle\bar{q}q\rangle F_{\mu\nu}.
\end{array}
\end{equation}
Note that $\chi$ has the dimension of GeV$^{-2}$, while 
$\kappa$ and $\xi$ are dimensionless.

\begin{figure}[p]
\centerline{\psfig{file=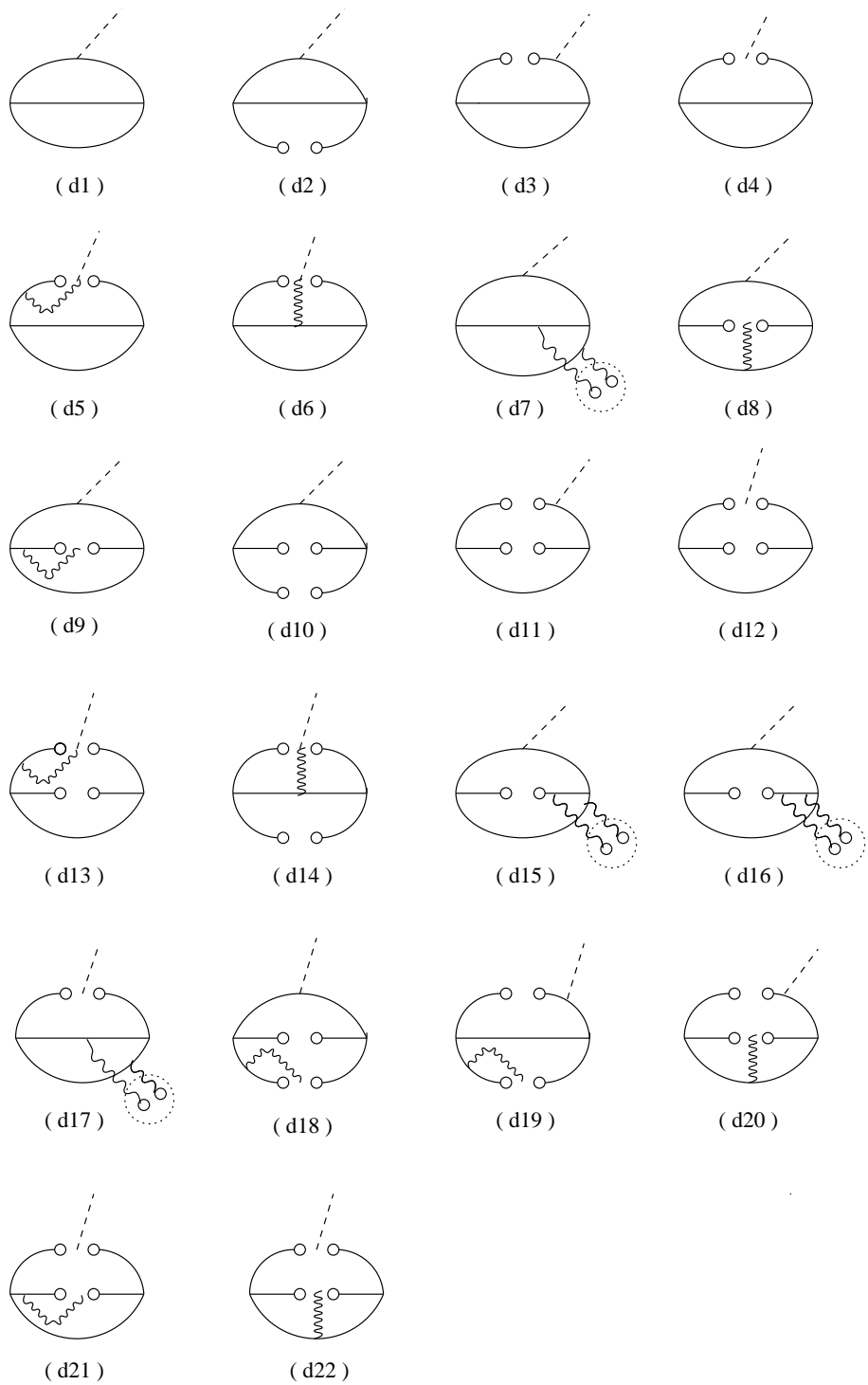,width=12cm}}
\vspace{1cm}
\caption{Diagrams considered for the decuplet baryon magnetic moments.}
\label{xmag}
\end{figure}

\begin{figure}[p]
\centerline{\psfig{file=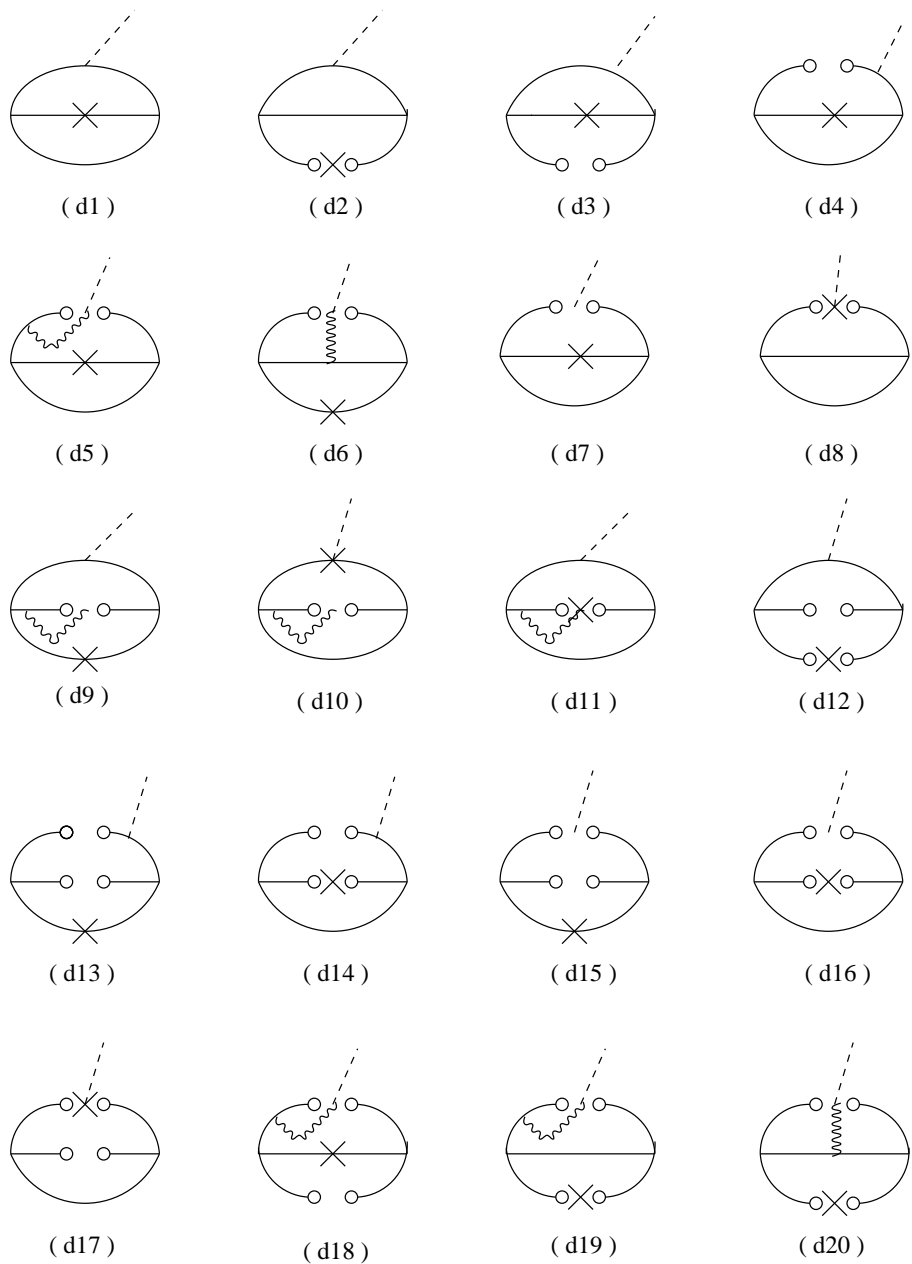,width=12cm}}
\vspace{1cm}
\caption{Diagrams considered for the strange quark mass corrections
to the decuplet baryon magnetic moments.}
\label{xmagm}
\end{figure}

The calculation proceeds by
substituting the quark propagator into the master formulae, 
keeping terms to first order in the external field and in the strange 
quark mass. 
Terms up to dimension 8 are considered.
The various combinations can be represented by diagrams.
Fig.~\ref{xmag} shows the basic diagrams considered
for the decuplet baryon magnetic moments.
Fig.~\ref{xmagm} 
shows the diagrams considered for the strange quark mass corrections.
Note that each diagram is only generic. All possible color permutations 
are understood. 
Numerous tensor structures emerge from the calculations.
Upon ordering the gamma matrices in the same order as in the 
phenomenological side, 18 invariant functions are obtained at 
the corresponding tensor structures.
By equating them with those in Eq.~(\ref{phen4}), 
QCD sum rules are constructed. 
These invariant functions can be classified by the chirality of the 
vacuum condensates they contain.
Eight of them , denoted by $\mbox{WE}_i$, 
involve only dimension-even condensates, 
thus we call the corresponding sum rules chiral-even. 
The other eight, denoted by $\mbox{WO}_i$, 
involve only dimension-odd condensates,
and we call the corresponding sum rules chiral-odd. 
Note that previous works such as Refs.~\cite{Ioffe84,Chiu86} use
the chirality of the tensor structures to refer to the sum rules.
The two are opposite.

To keep the presentation smooth,
the complete set of sum rules (a total of 160 for the decuplet family) 
obtained in this work are given in the Appendix in a highly condensed form.
As it turns out, the  validity of a particular sum rule depends on the 
input parameter set. Sum rules that are valid for one set may become 
invalid for another, and {\it vice versa}.
For this reason, it is useful to present all of the sum rules.
Another benefit is that it provides a basis for 
other authors to check the calculation. Sufficient detail is given in 
this work for that purpose.

The various symbols in the sum rules are explained in the following.
The condensate parameters are denoted by
\begin{equation}
a=-(2\pi)^2\,\langle\bar{u}u\rangle,
\hspace{2mm}
b=\langle g^2_c\, G^2\rangle, 
\hspace{2mm}
\langle\bar{u}g_c\sigma\cdot G u\rangle=-m_0^2\,\langle\bar{u}u\rangle.
\end{equation}
The re-scaled current coupling
\begin{equation}
\tilde{\lambda}_B=(2\pi)^2\lambda_B. 
\end{equation}
The quark charge factors $e_q$ are given in units of electric charge
\begin{equation}
e_u=2/3, 
\hspace{4mm}
e_d=-1/3, 
\hspace{4mm}
e_s=-1/3. 
\end{equation}
Note that we choose to keep the quark charge factors explicit in the 
sum rules. The advantage is that it can facilitate the study of quark 
effective magnetic moments.
The parameters $f$ and $\phi$ account for 
the flavor symmetry breaking of the strange quark in the 
condensates and susceptibilities:
\begin{equation}
f={ \langle\bar{s}s\rangle \over \langle\bar{u}u\rangle }
={ \langle\bar{s}g_c\sigma\cdot G s\rangle \over
   \langle\bar{u}g_c\sigma\cdot G u\rangle },
\hspace{4mm}
\phi={ \chi_s \over \chi }={ \kappa_s \over \kappa }={ \xi_s \over \xi }.
\end{equation}
The four-quark condensate is parameterized by the 
factorization approximation 
\begin{equation}
\langle\bar{u}u\bar{u}u\rangle=\kappa_v\langle\bar{u}u\rangle^2, 
\end{equation}
and we will investigate its possible violation via the parameter $\kappa_v$.
The anomalous dimension corrections of the currents and various operators 
are taken into account in the leading logarithmic approximation via the factor 
\begin{equation}
L^\gamma=\left[{\alpha_s(\mu^2) \over \alpha_s(M^2)}\right]^\gamma
=\left[{\ln(M^2/\Lambda_{QCD}^2) \over \ln(\mu^2/\Lambda_{QCD}^2)}
\right]^\gamma,
\end{equation}
where $\mu=500$ MeV is the
renormalization scale and $\Lambda_{QCD}$ is the QCD scale parameter.  
As usual, the excited state contributions are modeled using terms 
on the OPE side surviving $M^2\rightarrow \infty$ under the assumption
of duality, and are represented by the factors 
\begin{equation}
E_n(x)=1-e^{-x}\sum_n{x^n \over n!}, \hspace{3mm} x=w_B^2/M_B^2,
\end{equation}
where $w_B$ is an effective continuum threshold. Note that $w_B$
is in principle different for different sum rules and we
will treat it as a free parameter in the the analysis.

The coefficients for other members of the decuplet family 
can be obtained by appropriate replacements of quark contents.
They are:
\begin{enumerate}
\item
for $\Delta^{++}$, replace s quark by u quark in $\Omega^-$,
\item
$\Delta^+$: replace s quark by d quark in ${\Sigma^*}^+$,
\item
for $\Delta^0$: replace s quark by d quark in ${\Xi^*}^0$,
\item
for $\Delta^-$, replace s quark by d quark in $\Omega^-$,
\item
for ${\Sigma^*}^-$, replace u quark by d quark in ${\Sigma^*}^+$,
\item
for ${\Xi^*}^-$, replace u quark by d quark in ${\Xi^*}^0$.
\end{enumerate}
Here the conversions between u and d quarks are achieved 
by simply switching their charge factors $e_u$ and $e_d$.
The conversions from s quark to u or d quarks involve setting  
$m_s=0$, $f=\phi=1$, in addition to the switching of charge factors.

Furthermore, in the course of collecting the coefficients
for the four selected members 
${\Sigma^*}^+$, ${\Sigma^*}^0$, ${\Xi^*}^0$, $\Omega^-$,
we discovered some relations among them that allow one to 
write down one set of $c_i$ staring from another.
The relations are given as follows.
\begin{enumerate}
\item
From ${\Sigma^*}^+$ to ${\Sigma^*}^0$:
simply replace every occurrence of $e_u$ by $(e_u + e_d)/2$.
\item
From ${\Xi^*}^0$ to $\Omega^-$
involves converting the u quark to s quark.
This is achieved by collapsing each coefficient into a single term
that has the maximum  number of $e_s$, $f$, $\phi$ in that coefficient.
The numerical factor of it is the sum of the numerical factors in front 
each of the terms in the coefficient. For example, $(2e_s+e_u)$ goes to $3e_s$,
$(2 e_s f \phi + e_u)$ goes to $3e_s f \phi$,
$(2e_s f - 3e_s - 3e_u f + e_u)$ goes to $-3e_s f$,  etc..
\end{enumerate}
These relations were also used as consistency checks of the calculation.

From the above discussions, we see that it is possible to write down 
the coefficients for all 
other members of the decuplet family starting just from those 
for ${\Sigma^*}^+$ and ${\Xi^*}^0$.
In the sum rule from  WE$_1$ in Eq.~(\ref{we1}),
the complete sets of $c_i$ are given for the four selected members 
${\Sigma^*}^+$, ${\Sigma^*}^0$, ${\Xi^*}^0$ and $\Omega^-$. 
They are intended as 
examples for the reader to get familiar with the relations.
The rest of the sum rules are presented with $c_i$ 
only given for ${\Sigma^*}^+$ and ${\Xi^*}^0$.

Finally, let us point out some exact relations among the 
OPE sides of the sum rules:   
\begin{equation}
\mbox{OPE}_{\scriptscriptstyle \Delta^+}=
{1\over 2} \, \mbox{OPE}_{\scriptscriptstyle \Delta^{++}},
\label{exact1}
\end{equation}
\begin{equation}
\mbox{OPE}_{\scriptscriptstyle \Delta^0}=0,
\label{exact2}
\end{equation}
\begin{equation}
\mbox{OPE}_{\scriptscriptstyle \Delta^-}=
-\, \mbox{OPE}_{\scriptscriptstyle \Delta^+}.
\label{exact3}
\end{equation}
\begin{equation}
\mbox{OPE}_{\scriptscriptstyle {\Sigma^*}^0}={1\over 2} \, 
( \mbox{OPE}_{\scriptscriptstyle {\Sigma^*}^+} 
+ \mbox{OPE}_{\scriptscriptstyle {\Sigma^*}^-} ).
\label{exact4}
\end{equation}
These results are consequences of symmetries in the correlation functions.
As an example, let us examine Eq.~(\ref{exact1}).
For a given diagram,
the master formula for $\Delta^{++}$ can be written as 
$2(e_uC_1 + e_uC_2)={4\over 3}(C_1+C_2)$ 
where $C_1$ has the trace dependence, while $C_2$ not.
On the other hand, the master formula for $\Delta^{+}$ can be written as
${2\over 3}[(2e_u+e_d)C_1 + (2e_u+e_d)C_2]={2\over 3}(C_1+C_2)$,
hence the factor of 2.
The key here is: a) each term is proportional to a quark charge factor;
b) SU(2) flavor symmetry in u and d quarks;
c) it is the same $C_1$ and $C_2$ that appear in both cases.
The argument can be generalized to any diagrams,  only
with $C_1$ and $C_2$ different from diagram to diagram.
Thus the factor of 2 will survive, regardless of the number 
of diagrams considered.
One can argue for the rest of the relations by the same token.
The above results have been explicitly verified using the 
calculated coefficients in the sum rules.
They also provided a set of highly non-trivial checks of the calculation.
A number of hard-to-detect errors have been eliminated this way.

Now let us consider the phenomenological side of Eq.~(\ref{exact1}).
Since the continuum is modeled using terms on the OPE side,
the continuum contributions also differ by a factor of 2.
Assuming the transitions, which are modeled by a constant, 
also differ by a factor of 2, then Eq.~(\ref{exact1}) 
can be extended to the magnetic moments.
This assumption was confirmed by numerical analysis.
The same is true for Eq.~(\ref{exact2}) and Eq.~(\ref{exact3}).
The situation for Eq.~(\ref{exact4}) is a little different. 
The convergence properties may change when two OPE 
series are added up.
Numerical analysis confirmed that fewer sum rules are valid 
for ${\Sigma^*}^0$ than for ${\Sigma^*}^+$ and ${\Sigma^*}^-$.

\section{Monte-Carlo Analysis}
\label{ana}

To analyze the sum rules, we use a Monte-Carlo based
procedure recently developed in Ref.~\cite{Derek96}.
The basic steps are as follows.
First, the uncertainties in the QCD input parameters are assigned.
Then, randomly-selected, Gaussianly-distributed sets for these 
uncertainties are generated,
from which an uncertainty distribution in the OPE, 
$\sigma^2_{\scriptscriptstyle OPE}(M_j)$ 
where $M_j$ are evenly distributed 
points in the desired Borel window,  can
be constructed.  
Next, a $\chi^2$ minimization is applied to the sum
rule by adjusting the phenomenological fit parameters. 
Note that the uncertainties in the OPE are not uniform throughout the 
Borel window. They are larger at the lower end where uncertainties in the 
higher-dimensional condensates dominate.
Thus, it is crucial that the appropriate weight is used in the
calculation of $\chi^2$.
For the OPE obtained from the k'th set of QCD parameters, the 
$\chi^2$ per degree of freedom is 
\begin{equation}
{\chi^2_k\over N_{DF}}={1\over n_B-n_p} 
\sum^{n_{\scriptscriptstyle B}}_{j=1} 
{ [\Pi^{\scriptscriptstyle OPE}_k(M_j)
-\Pi^{\scriptscriptstyle Phen}_k(M_j;\lambda_k,m_k,w_k)]^2
\over \sigma^2_{\scriptscriptstyle OPE}(M_j) },
\end{equation}
where $n_p$ is the number of phenomenological search parameters,
and $\Pi^{\scriptscriptstyle Phen}$ 
denotes the phenomenological representation.
In practice, $n_B$=51 points were used along the Borel axis.
The procedure is repeated for many QCD parameter sets, 
resulting in distributions for phenomenological fit parameters, 
from which errors are derived.
Usually, 200 such configurations are sufficient for getting stable
results. We generally select 1000 sets which help resolve more subtle
correlations among the QCD parameters and the phenomenological fit
parameters.

The Borel window over which the two sides
of a sum rule are matched is determined by the following two criteria.
First, {\em OPE convergence}: the highest-dimension-operators
contribute no more than 10\% to the QCD side.
Second, {\em ground-state dominance}: excited
state contributions should not exceed more than 50\% of the 
phenomenological side.
The first criterion effectively establishes a lower limit, 
the second an upper limit.
Those sum rules which do not have a Borel window under these 
criteria are considered invalid.

\subsection{QCD Input Parameters}
\label{input}

The QCD input parameters and their uncertainty assignments are 
given as follows.  The condensates are taken as 
\begin{equation}
a=0.52\pm0.05 \;GeV^3, \;\;
b=1.2\pm0.6 \;GeV^4, \;\;
m^2_0=0.72\pm0.08 \;GeV^2.
\end{equation}
For the factorization violation parameter, we use 
\begin{equation}
\kappa_v=2\pm 1 \;\;\mbox{and}\;\; 1\leq \kappa_v \leq 4.
\end{equation}
The QCD scale parameter is restricted to $\Lambda_{QCD}$=0.15$\pm$0.04
GeV.
The vacuum susceptibilities have been estimated in 
studies of nucleon magnetic moments~\cite{Ioffe84,Balitsky83,Chiu86},
but the values vary in a wide range depending on the method used.
Here we take some median values with 50\% uncertainties:
\begin{equation}
\chi=-6.0\pm 3.0\; GeV^{-2}
\;\; \mbox{and}\;\; 0 \; GeV^{-2}\leq \chi \leq -10 \;GeV^{-2},
\end{equation}
and
\begin{equation}
\kappa=0.75\pm 0.38,\;\; \xi=-1.5\pm 0.75.
\end{equation}
Note that $\chi$ is almost an order of magnitude 
larger than $\kappa$ and $\xi$, and is the most important of the three.
The strange quark parameters are placed at~\cite{Pasupathy86,Lee97a} 
\begin{equation}
m_s=0.15\pm 0.02 \;GeV, \;\;
f=0.83\pm0.05, \;\;
\phi=0.60\pm0.05.
\end{equation}
These uncertainties are assigned conservatively and in accord with the
state-of-the-art in the literature. 
While some may argue that some
values are better known, others may find that the errors are
underestimated.  In any event, one will learn how the uncertainties in
the QCD parameters are mapped into uncertainties in the
phenomenological fit parameters. In the numerical analysis below,
we will also examine how the spectral parameters depend on different 
uncertainty assignments in these input parameters.

\subsection{Search Procedure}
\label{search}

To extract the magnetic moments, a two-stage fit was performed. 
First, the corresponding chiral-odd mass sum rule, as obtained 
previously in  Ref.~\cite{Lee97a}, was fitted to get the mass
$M_B$, the coupling $\tilde{\lambda}_B^2$ and the continuum threshold
$w_1$.  Then, $M_B$ and $\tilde{\lambda}_B^2$ were used in the 
magnetic moment sum rule for a three-parameter fit:
the transition strength $A$, the continuum threshold $w_2$, 
and the magnetic moment $\mu_{\scriptscriptstyle B}$.
Note that $w_1$ and $w_2$ are not necessarily the same.
We impose a physical constraint on both $w_1$ and $w_2$ 
requiring that they are larger than the mass,
and discard QCD parameter sets that do not satisfy this condition.
In the actual analysis of the sum rules,
however, we found that a full search was not always successful. 
In such cases, the search algorithm consistently returned $w_2$
either zero or smaller than $M_B$. This signals insufficient information
in the OPE to completely resolve the spectral parameters.
To proceed, we fixed $w_2$ at $w_1$, which is a
commonly-adopted choice in the literature, 
and searched for $A$ and $\mu_{\scriptscriptstyle B}$.
The two-stage fit incorporates the uncertainties from 
the two-point functions in a correlated fashion into
the three-point functions, and  represents a more realistic scenario.

To illustrate how well a sum rule works, we first cast it into 
the subtracted form,
\begin{equation}
\Pi_S=\tilde{\lambda}^2_B\, \mu_{\scriptscriptstyle B}\, e^{-M^2_B/M^2},
\end{equation}
then plot the logarithm of the absolute value of the two sides
against the inverse of $M^2$.
In this way, the right-hand side will appear as a straight line
whose slope is $-M_B^2$ and whose intercept with the y-axis gives
some measure of the coupling strength and the magnetic moment.
The linearity (or deviation from it) of the left-hand side gives 
an indication of OPE convergence, and information 
on the continuum model and the transitions.
The two sides are expected to match for a good sum rule.
This way of matching the sum rules is similar to looking for a
`plateau' as a function of Borel mass in the conventional analysis, 
but has the advantage of not restricting the analysis regime in Borel 
space to the valid regimes common to {\it both} two-point and 
three-point correlation functions.

\section{Results and Discussions}
\label{res}

\begin{figure}[p]
\centerline{\psfig{file=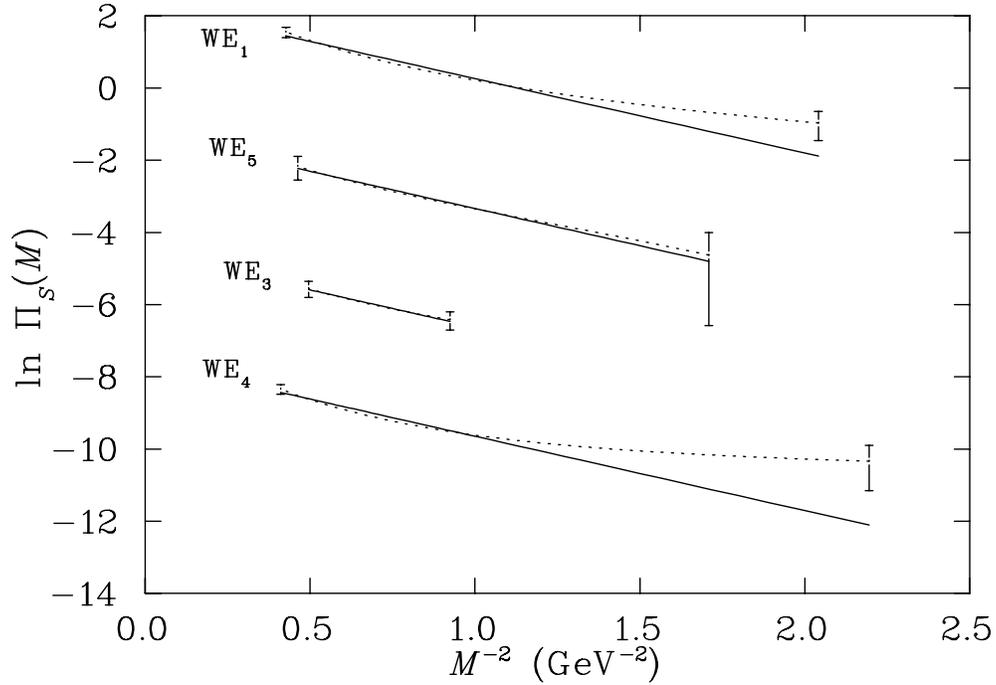,height=9cm,width=13cm,angle=90}}
\vspace{1cm}
\caption{Overlap plots of the valid 
QCD sum rules for the $\Delta^{++}$ magnetic moment.
Each sum rule is searched independently.
The solid line corresponds to the ground state contribution, 
the dotted line the rest of the contributions 
(OPE minus continuum minus transition).
The error bars are only shown at the two ends for clarity.
From top down, the sum rules are arranged by magnitudes of 
$\mu_{\scriptscriptstyle B}$ extracted from them.
For better viewing, the curves for each sum rule are shifted downward
by 3 units relatively to the previous one.}
\label{rhslhs-delpp}
\end{figure}

\begin{figure}[p]
\centerline{\psfig{file=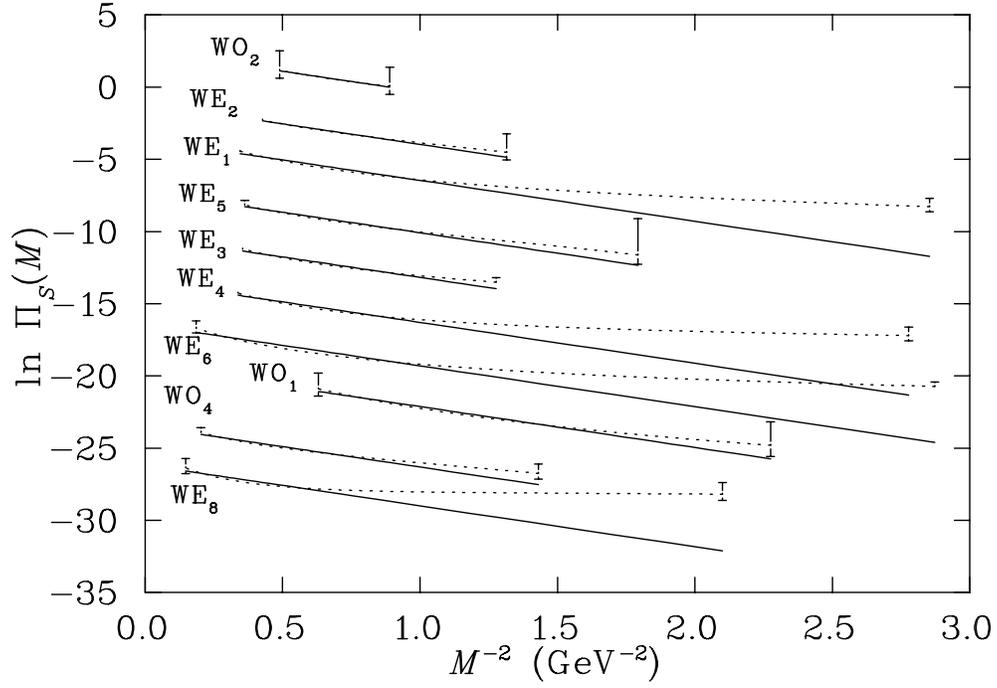,height=9cm,width=13cm,angle=90}}
\vspace{1cm}
\caption{Similar to Fig.~\protect\ref{rhslhs-delpp}, but for
$\Omega^{-}$.}
\label{rhslhs-omeg}
\end{figure}

\begin{figure}[p]
\centerline{\psfig{file=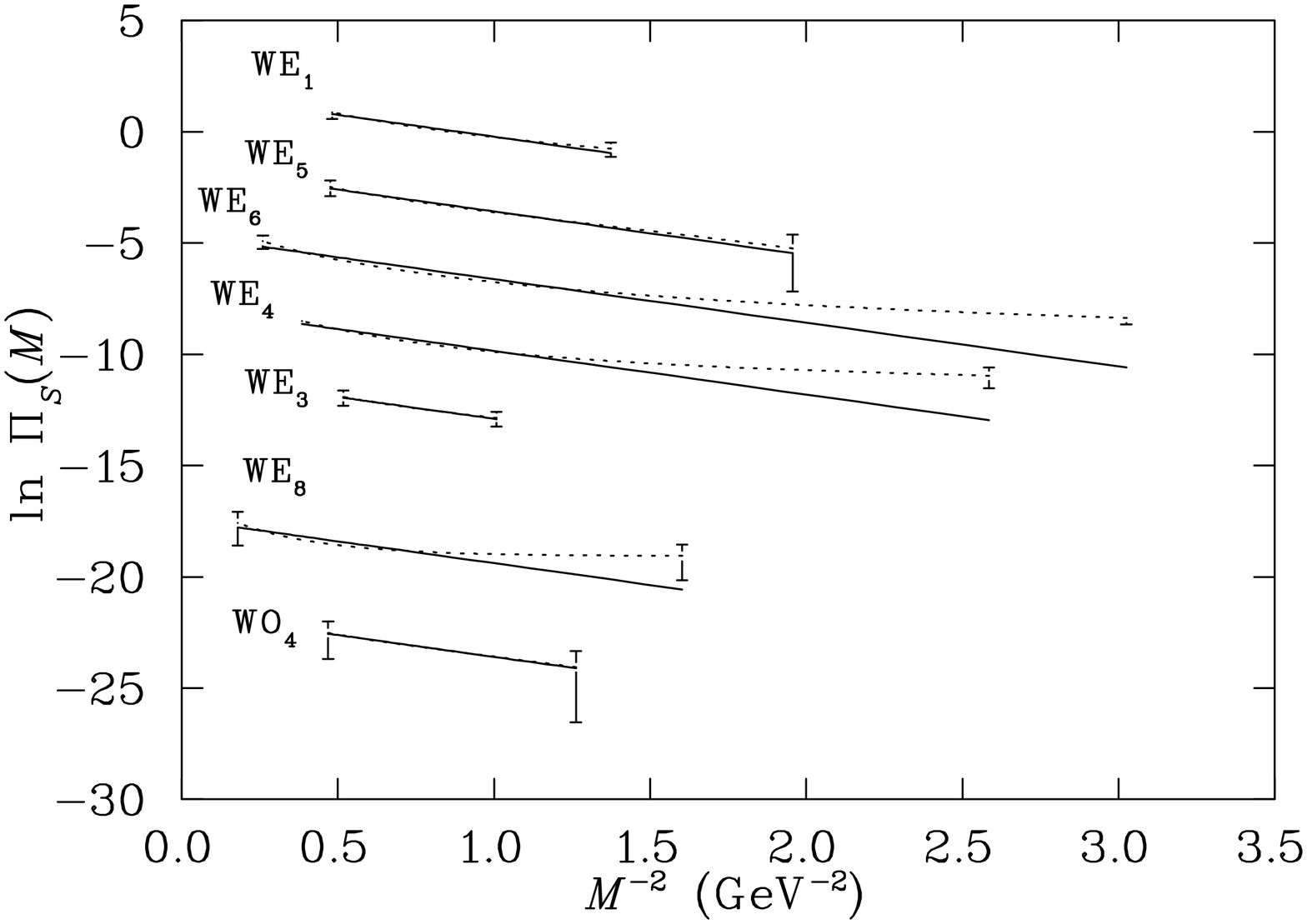,height=9cm,width=13cm,angle=90}}
\vspace{1cm}
\caption{Similar to Fig.~\protect\ref{rhslhs-delpp}, but for
${\Sigma^*}^{+}$.}
\label{rhslhs-sigp}
\end{figure}

\begin{figure}[p]
\centerline{\psfig{file=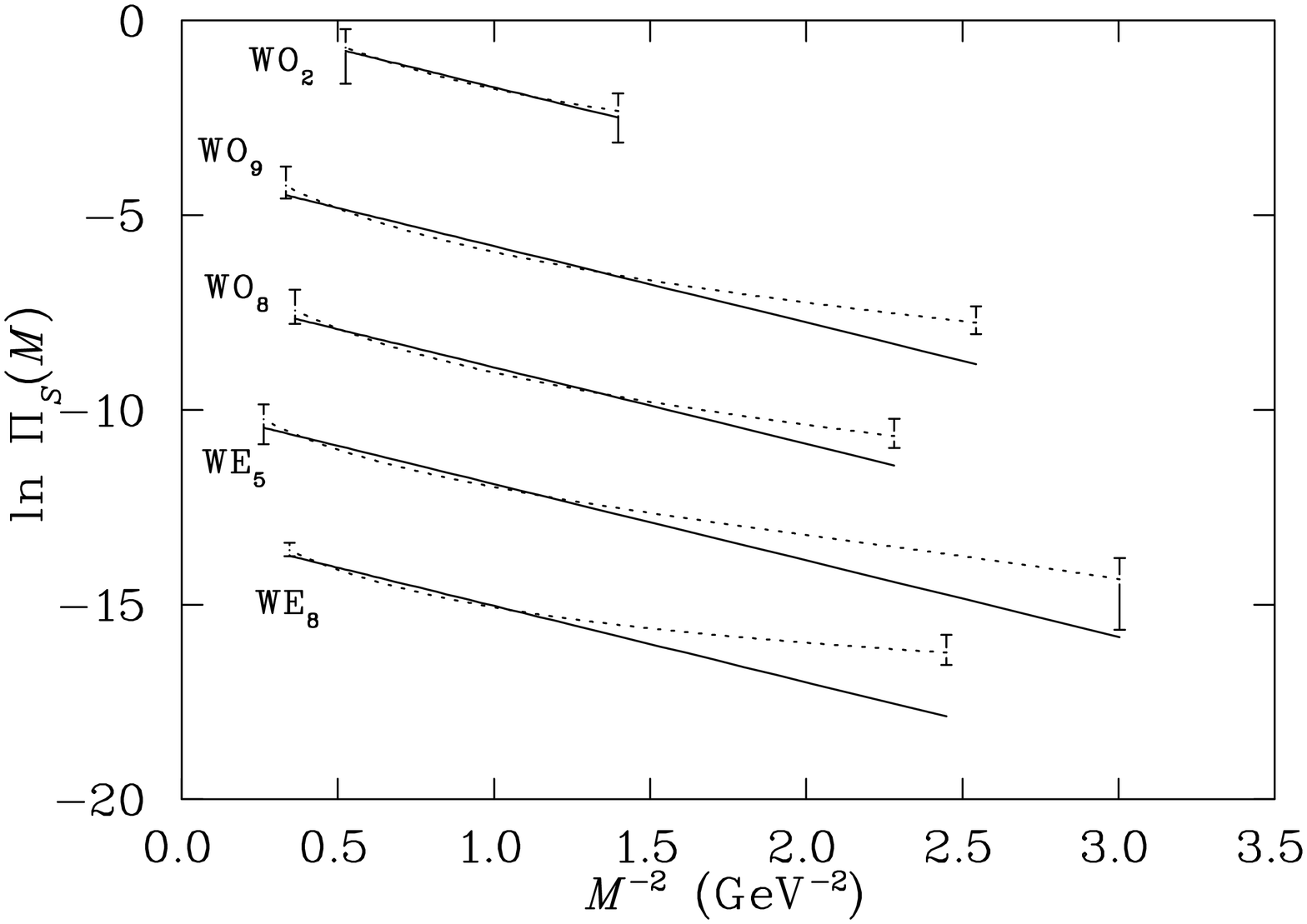,height=9cm,width=13cm,angle=90}}
\vspace{1cm}
\caption{Similar to Fig.~\protect\ref{rhslhs-delpp}, but for
${\Sigma^*}^{0}$.}
\label{rhslhs-sig0}
\end{figure}

\begin{figure}[p]
\centerline{\psfig{file=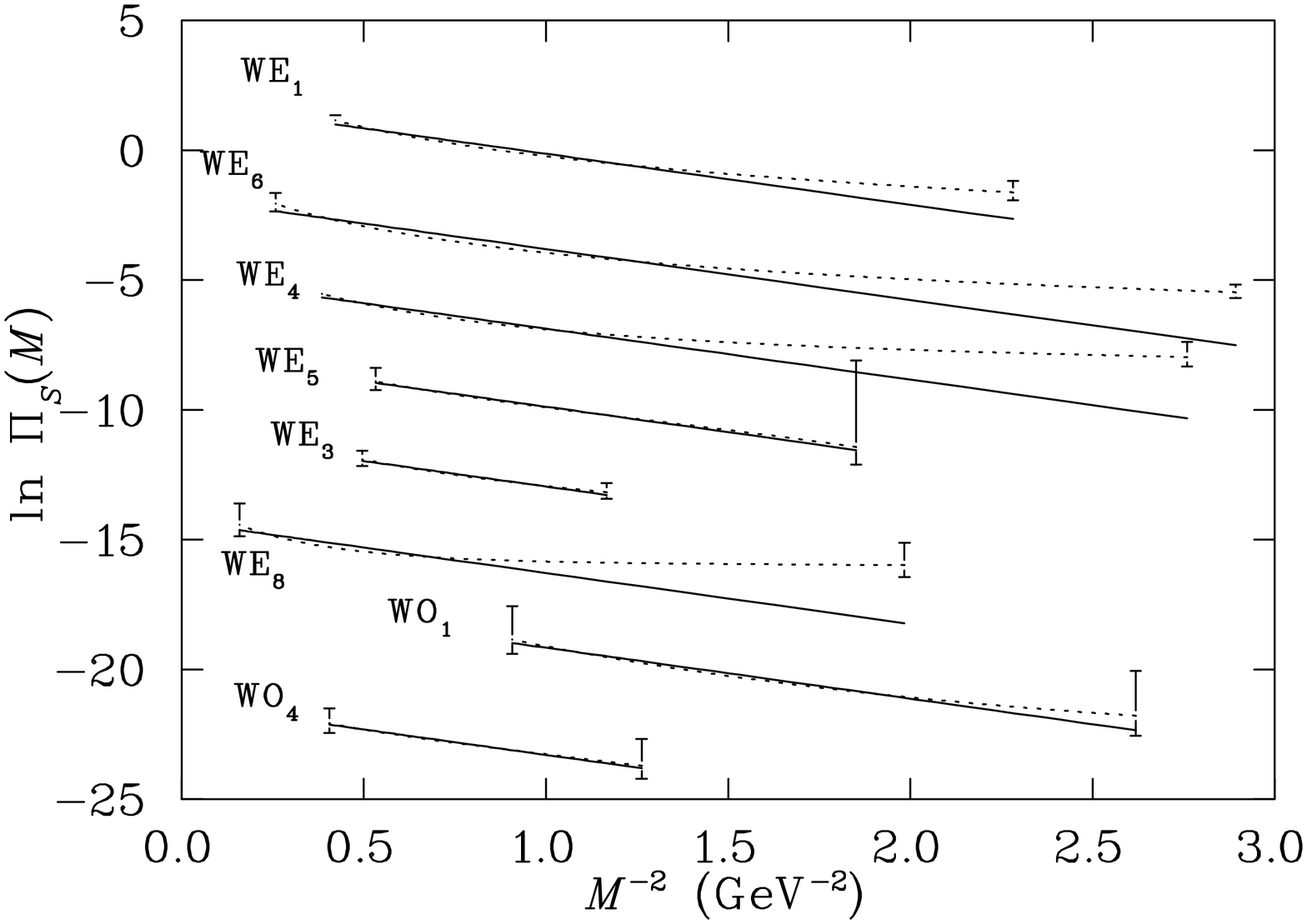,height=9cm,width=13cm,angle=90}}
\vspace{1cm}
\caption{Similar to Fig.~\protect\ref{rhslhs-delpp}, but for
${\Sigma^*}^{-}$.}
\label{rhslhs-sigm}
\end{figure}

\begin{figure}[p]
\centerline{\psfig{file=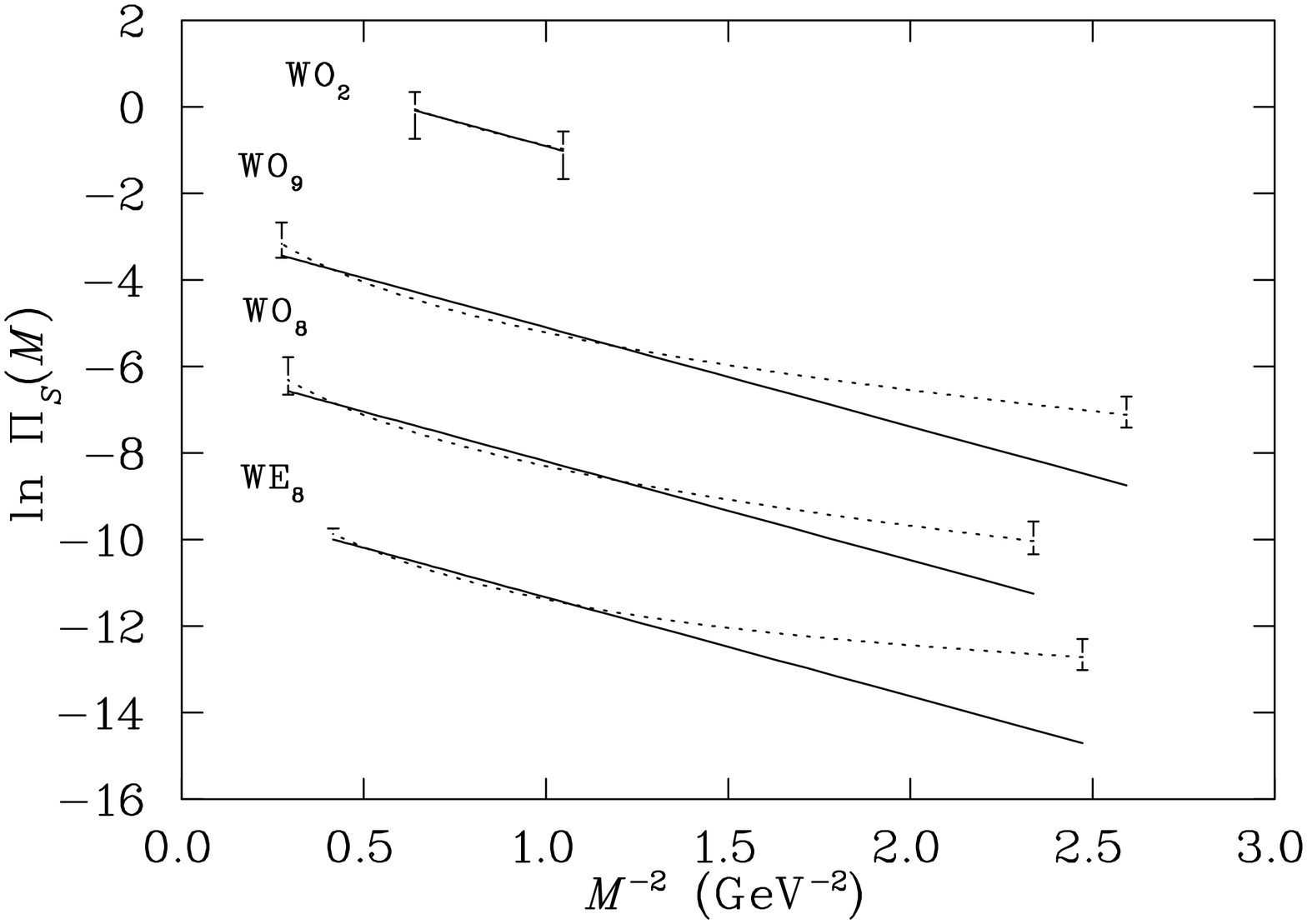,height=9cm,width=13cm,angle=90}}
\vspace{1cm}
\caption{Similar to Fig.~\protect\ref{rhslhs-delpp}, but for
${\Xi^*}^{0}$.}
\label{rhslhs-xi0}
\end{figure}

\begin{figure}[p]
\centerline{\psfig{file=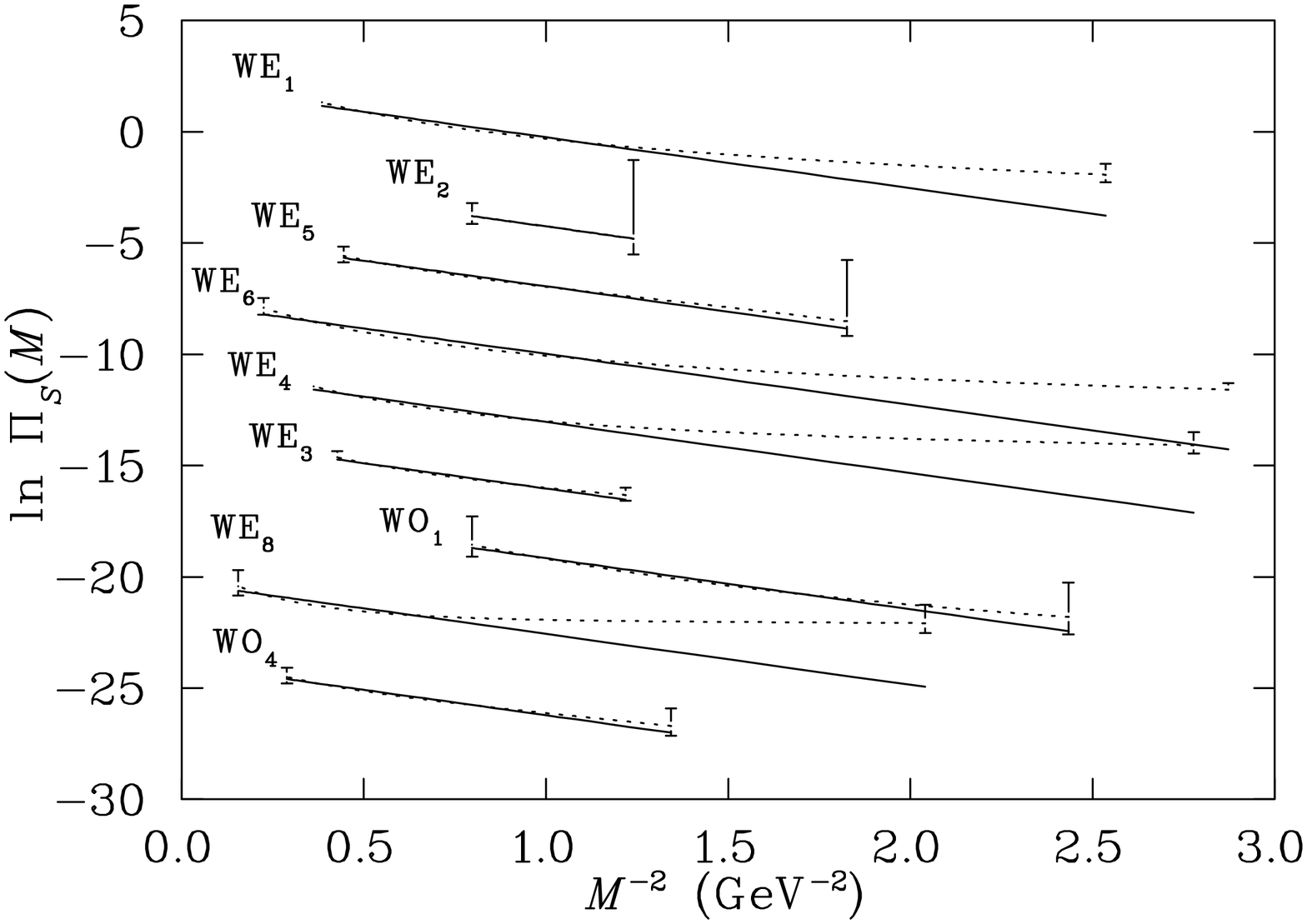,height=9cm,width=13cm,angle=90}}
\vspace{1cm}
\caption{Similar to Fig.~\protect\ref{rhslhs-delpp}, but for
${\Xi^*}^{-}$.}
\label{rhslhs-xim}
\end{figure}

We have analyzed all of the sum rules for the entire decuplet family.
We confirmed the three relations among magnetic moments as extended from 
Eqs.~(\ref{exact1}) to~(\ref{exact3}). 
So we will only present results for seven members.
Valid sum rules were identified using the criteria discussed earlier. 
The results are given in three tables: Tables~\ref{tabdo} to~\ref{tabxi}. 
The corresponding overlap plots are given in 
seven figures: Figs.~\ref{rhslhs-delpp} to~\ref{rhslhs-xim}.
These plots show how well a sum rule performs in the entire Borel region.       
Such information is absent in the tables.
From the results, the following observations are in order.

In general, more chiral-even sum rules are valid than chiral-odd ones.
This is consistent with previous findings for the octet 
baryon magnetic moments.  It was argued in Ref.~\cite{Ioffe84} 
that the interval of dimensions (not counting the dimension of 
$F_{\mu\nu}$) in the chiral-even sum rules (0 to 8) is 
larger than that in the chiral-odd sum rules (1 to 7).
Indeed, more chiral-even sum rules 
(WE$_2$, WE$_4$, WE$_5$, WE$_6$, WE$_8$, WE$_9$) 
have power corrections up to $1/M^4$,
than chiral-odd ones (WO$_2$, WO$_4$, WO$_8$, WO$_9$). 
Because of the additional terms in the OPE series, these sum rules 
are expected to be more reliable than the other sum rules.
The situation here is almost opposite to   that for the two-point 
functions~\cite{Lee97a}. It was pointed out in Ref.~\cite{Jin97} that chiral-odd 
sum rules are more reliable than chiral-even sum rules for baryon 
two-point functions.
The reason could be traced to the fact that even and odd parity excited 
states contribute with different signs.
In the three-point functions, however, the statement is no longer 
valid due to the appearance of transitions and vacuum susceptibilities.
Therefore, caution should be used when applying the 
chirality argument to determine the reliability of a sum rule in 
three-point functions.
In addition, numerical analysis showed that the sum rules from 
WE$_1$, WE$_3$, WO$_1$ are valid for the standard input parameter set,
despite the absence of $1/M^4$ terms. 
We have varied the central values 
of the input parameters and discovered that sum rules that were valid 
for one set of input parameters became invalid for another, 
and {\it vice versa}.
Thus the situation with three-point functions is more complicated.
Our experience is that each sum rule should be examined individually
in order to find out its reliability.

It turns out that for most of the valid sum rules, a full search was
unsuccessful, except for three sum rules: WE$_5$, WE$_6$ and WE$_8$ 
for ${\Sigma^*}^{+}$.
Of the three, only WE$_6$ returned a continuum threshold with 
reasonable error.
The other two returned  it with large errors.
The large errors indicate that the sum rules are not very stable:
they contain barely enough information to completely resolve 
the spectral parameters..
The important point is that the results with the continuum threshold
searched or not are almost the same. This suggests that fixing it 
to that of the corresponding two-point function seems a good approximation.

It is gratifying to observe that the valid sum rules for most members
give consistent predictions for the magnetic moments in terms of the sign,
except for ${\Sigma^*}^{0}$ and ${\Xi^*}^0$ whose magnitudes are small.
The magnitudes for the magnetic moments are consistent within errors 
for the most part, with only a few exceptions. 
The performances of the sum rules are quite different within each member.
This is best displayed in the overlap plots.
In some sum rules, the overlap is poor,
as evidenced by the deviation from linearity (dotted lines).
It signals poor OPE convergence in these sum rules.
As expected, the deviation is more severe in the lower end of 
the Borel region where nonperturbative physics dominates.
These sum rules will more likely suffer from uncertainties associated with
the selection of the Borel window.
As a result, the spectral parameters extracted from them are less reliable.
One way to alleviate the problem is to increase the lower end of the 
Borel window to values where the overlap is good, even to extend the 
upper end to ensure the existence of a window.
This was not attempted in this work because we feel
the results obtained this way are somewhat misleading.
The reason is that the sum rules in these windows will be dominated 
mostly by perturbative physics. 
It is common knowledge that if one goes deep enough into the Borel space, 
one can always find a match in a QCD sum rule. 
But such practice is against the philosophy of the QCD sum rule approach, 
which relies upon the power corrections to resolve the spectral properties.
Therefore, some standard is necessary to emphasize such physics,
and we feel the 10\%-50\% criteria adopted here are a reasonable choice.

Based on the quality of the overlap, 
the broadness of the Borel window and its reach into the lower end,
the size of the continuum contribution, and the standard QCD input 
parameter set, we designate one sum rule 
for each member as the most favorable. They are 
WE$_5$ for $\Delta^{++}$, 
WE$_5$ for ${\Sigma^*}^{+}$,
WO$_8$ for ${\Sigma^*}^{0}$,
WE$_5$ for ${\Sigma^*}^{-}$,
WO$_8$ for ${\Xi^*}^0$,
WE$_5$ for ${\Xi^*}^-$,
and WE$_5$ for $\Omega^{-}$.
The selection is undoubtedly subjective.
The reader may find a different set that have equal or comparable performance.
We want to stress that such a selection depends on the QCD 
input parameters. It is possible that the ones selected here become invalid   
for a different set of input parameters, in which case a new set should 
be selected.

Relatively large errors were found in the valid sum rules using the 
standard QCD input parameter set: 
from 50\% to 100\% in the magnetic moments. 
But in most cases, the sign and order of magnitudes are 
unambiguously predicted when compared to the measured values.
The situation is similar to a previous finding on $g_A$~\cite{Lee97}.
To gain some idea on how the uncertainties depend on the input,
we also analyzed the sum rules by adjusting the error estimates
individually.
We found large sensitivities to the quark condensate magnetic 
susceptibility $\chi$.
In fact, most of the errors came from the uncertainties in $\chi$.
We also tried with reduced error estimates on all
the QCD input parameters: 10\% relative errors uniformly.
It leads to about 30\% accuracy on the magnetic moments in 
the favorable sum rules.
Further improvement of the accuracy by 
reducing the errors in the input is beyond the capability of these sum rules
as the $\chi^2/N_{DF}$ becomes unacceptably large, signaling 
internal inconsistency of the sum rules.
For that purpose, one would have to resort to finding sum rules 
that have better convergence properties and depend less critically on 
the poorly-known $\chi$.

\begin{figure}[p]
\centerline{\psfig{file=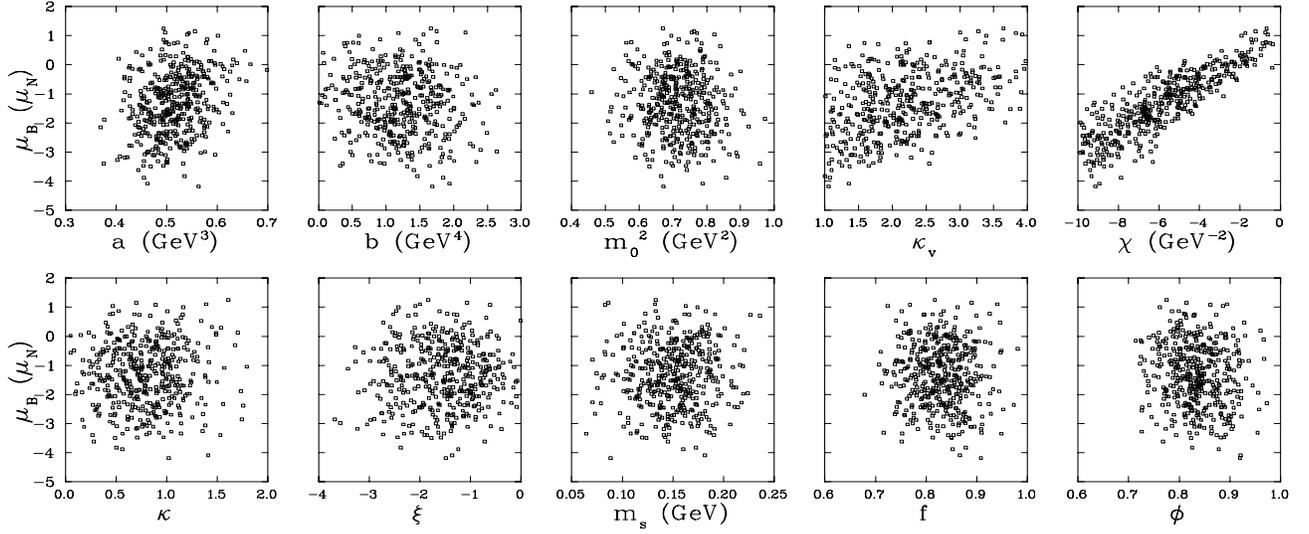,height=7cm,width=17cm,angle=90}}
\vspace{1cm}
\caption{Scatter plots showing correlations between the magnetic
moment of $\Omega^-$ and the standard QCD input parameters 
for the sum rule from WE$_5$. The result is drawn from 430 QCD parameters sets.}
\label{corr-omeg-we5}
\end{figure}

\begin{figure}[p]
\centerline{\psfig{file=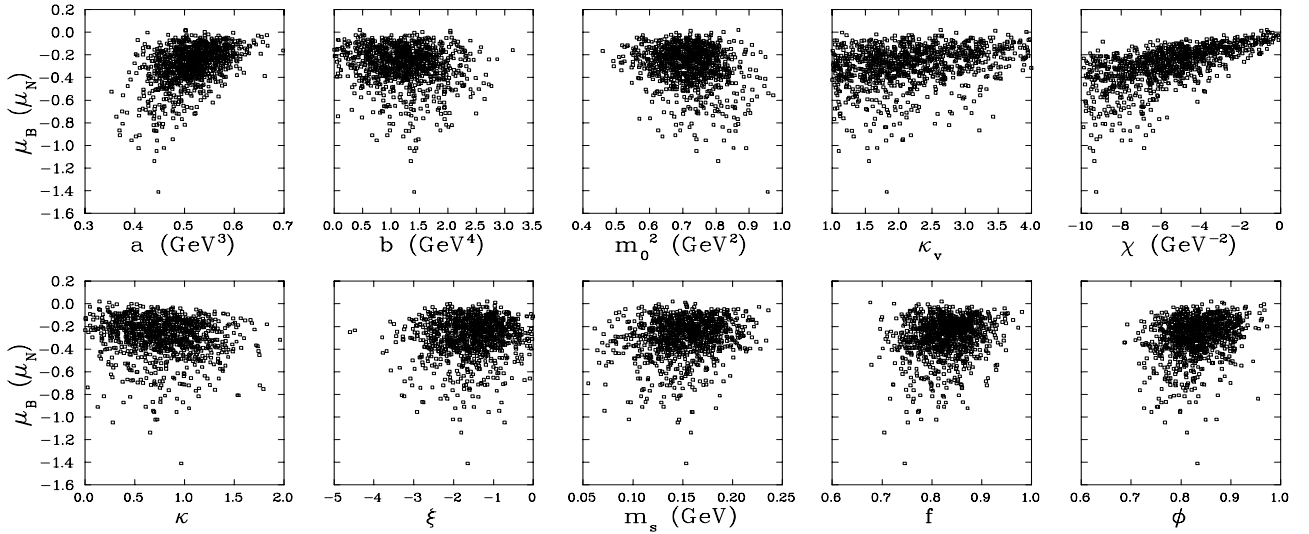,height=7cm,width=17cm,angle=90}}
\vspace{1cm}
\caption{Similar to Fig.~\protect\ref{corr-omeg-we5}, but for
${\Sigma^*}^{0}$ and the sum rule from WO$_8$.
The result is drawn from 1000 QCD parameters sets.}
\label{corr-sig0-wo8}
\end{figure}

To get a different perspective on how the spectral parameters depend 
on the input parameters, we study correlations among the parameters by
way of scatter plots. 
In the Monte-Carlo analysis, all the parameters are correlated. 
Therefore, one can study the correlations between any two parameters
by looking at their scatter plots.  Such plots are useful in
revealing how a particular sum rule resolves the spectral properties. 
We have examined numerous such plots. Here we focus on the 
favorable sum rules as selected earlier. 
To conserve space, we only give two examples.
Fig.~\ref{corr-omeg-we5} shows the scatter plot for correlations 
between $\Omega^-$ magnetic moment and the QCD input parameters for the 
sum rule from WE$_5$. Fig.~\ref{corr-sig0-wo8} shows a similar plot 
for ${\Sigma^*}^0$ and the sum rule from WO$_8$.
Perhaps the most interesting feature is the strong correlations with 
$\chi$ in both sum rules. 
This is the reason for the large sensitivities to this parameter 
as alluded to earlier. Precise determination of $\chi$ is crucial for 
keeping the uncertainties in the spectral parameters under control.
Other charged members (all use sum rules from WE$_5$)
display qualitatively the same patterns for parameters other than 
$\chi$ and the factorization violation parameter $\kappa_v$.
For $\chi$, positively-charged members ($\Delta^{++}$ and ${\Sigma^*}^+$) 
show negative correlations. The opposite is true for 
negatively-charged members ($\Omega^-$, ${\Sigma^*}^-$ and ${\Xi^*}^-$):
they show positive correlations with $\chi$.
The patterns for $\kappa_v$ essentially follow those for $\chi$, although
the correlations are weaker. 
The correlation patterns for ${\Xi^*}^0$ are qualitatively the same as 
those for ${\Sigma^*}^0$.

Table~\ref{comp} shows a comparison of the magnetic moments from 
various calculations and existing experimental data. 
The results with 10\% errors from the QCD sum rule method are used 
in the comparison. Note that the central values are slightly 
different from those in Tables~\ref{tabdo} to~\ref{tabxi} where 
conservative uncertainties were used.
The reason is that the resultant distributions vary with input 
errors and are not Gaussian in this case.
In such event the median and the average of the asymmetric errors 
are quoted. 
The QCDSR results are consistent with data, although the central 
value for $\Omega^-$ is slightly underestimated. 
We would like to point out that it is possible to reproduce the 
central value for $\Omega^-$ (using it as input) by fine-tuning of 
the susceptibility $\chi$ alone, given the sensitivity to this parameter 
and the large freedom at the present time on its value.
However, we feel that such an attempt is not very meaningful given the 
accuracy of the method. A more meaningful practice would be to 
re-analyze the octet baryon magnetic moments 
by the same method as employed here, and obtain a best fit on the
the susceptibilities using their accurately measured values,
then use them to predict the decuplet magnetic moments.
It would yield valuable information on these important quantities 
and on the consistency of the approach.
From the table, it is fair to say that the QCDSR approach is 
at least competitive with other calculations.
The results came about from a rather different perspective: the 
nonperturbative structure of the QCD vacuum.
The results from various calculations roughly agree, 
except for the charge-neutral resonances
$\Delta^{0}$, ${\Sigma^*}^{0}$, and ${\Xi^*}^0$ for which both the sign 
and the magnitude vary.
It would be helpful to have experimental information on the other members 
of the decuplet, although such measurements appear difficult.

\section{Conclusion}
\label{sumcon}

It has been demonstrated in this work 
that the magnetic moments of decuplet 
baryons can be successfully computed in the QCD sum rule approach.
A complete set of QCD sum rules are  derived using the 
external field technique.
They are analyzed extensively 
with a comprehensive Monte-Carlo based procedure 
which , in our opinion, provides the most realistic estimates 
of the uncertainties present in the approach.
 
Valid sum rules are identified using criteria established by 
OPE convergence and ground-state dominance. 
For each member, usually several sum rules are valid,
but not all of them perform equally well.
This was best displayed by the overlap plots.
Some have large deviations in the lower end of the Borel window,
signaling insufficient convergence in the OPE.
These sum rules are less reliable. 
Based on overall performance,
a favorable sum rule was selected for each member.
They are WE$_5$ for charged members, WO$_8$ for charge-neutral members.
We also found the following relations between the magnetic moments:
$\mu_{\scriptscriptstyle \Delta^+}=
{1\over 2} \, \mu_{\scriptscriptstyle \Delta^{++}}$,
$\mu_{\scriptscriptstyle \Delta^0}=0$,
and $\mu_{\scriptscriptstyle \Delta^-}=
-\, \mu_{\scriptscriptstyle \Delta^+}$, 
and approximately
$\mu_{\scriptscriptstyle {\Sigma^*}^0}={1\over 2}
(\mu_{\scriptscriptstyle {\Sigma^*}^r+}
+ \mu_{\scriptscriptstyle {\Sigma^*}^-} )$.
.

Using conservative estimates of the QCD input parameters,
the uncertainties in the extracted magnetic moments are found 
relatively large as compared to the two-point functions. 
We found that the results are sensitive to 
the quark condensate magnetic susceptibility $\chi$. 
In fact, most of the uncertainties could be attributed to $\chi$.
Better estimate of this parameter is clearly needed.
By varying the uncertainty estimates in the input parameters, 
we found that a 30\% accuracy can be achieved with the 
designated sum rules if the QCD input parameters could be 
determined to the 10\% accuracy level.

\acknowledgements
It is a pleasure to thank D.B. Leinweber for providing 
an original version of his Monte-Carlo
analysis program and for helpful discussions.
This work was supported in part by 
U.S. DOE under Grant DE-FG03-93DR-40774.

\appendix
\section{QCD Sum Rules for Magnetic Moments of Decuplet Baryons}
\label{qcdsr}

Here we give the complete set of QCD sum rules derived in this work.
Foe each member, there are 18 sum rules, 9 chiral-even, 9 chiral-odd.
It turns out that the sum rules from WE$_6$ and WE$_7$ are degenerate, 
so are those from WO$_4$ and WO$_5$.
So the number of independent sum rules is 16 for each member.
The total number for the entire decuplet family is 160.
They are given in the following in a highly compact form.
The explanation on how to obtain a sum rule for a 
particular member is discussed in the main text.

The sum rule from WE$_1$:
\begin{eqnarray}
& &
  c_1  L^{4/27} E_1 M^4
+ c_2  m_s \chi a L^{-12/27} E_0 M^2
+ c_3  b L^{4/27} 
+ c_4  \chi a^2 L^{12/27} 
+ (c_5+c_6)  m_s a L^{4/27} 
\nonumber \\ & &
+ (c_7+c_8)  a^2 L^{28/27} {1\over M^2}
+ c_9 \chi m^2_0 a^2 L^{-2/27} {1\over M^2}
+ c_{10}  m_s m^2_0 a L^{-10/27} {1\over M^2}
\nonumber \\ & &
={1\over 2} \tilde{\lambda}^2_B
\left( {\mu_{\scriptscriptstyle B}\over M^2} + A \right) e^{-M^2_B/M^2},
\label{we1}
\end{eqnarray}
where the coefficients for ${\Sigma^*}^+$ are:
\begin{equation}
\begin{array}{ll}
c_1={1\over 8}(e_s + 2e_u), & 
c_2={-7\over 18}(e_s f \phi + 2e_u), \\
c_3={-1\over 72}(e_s + 2e_u), &  
c_4={-1\over 9}(e_s f \phi + e_u f + e_u),\\
c_5={1\over 18}(-2e_s f + 9e_s + 9e_u f + 5e_u), & 
c_6={-1\over 18}(e_s f \phi + 2e_u)(7\kappa+\xi), \\
c_7={1\over 27}(-e_s f + 3e_s + 5e_u f - e_u)\kappa_v,& 
c_8={-1\over 54}(e_s f \phi + e_u f + e_u)(7\kappa+\xi),\\
c_9={7\over 216}(e_s f \phi + e_u f + e_u),&
c_{10}={-5\over 72}(e_s + e_u f + e_u),
\end{array}
\end{equation}
for ${\Sigma^*}^0$:
\begin{equation}
\begin{array}{ll}
c_1={1\over 8}(e_s + e_u + e_d), & 
c_2={-7\over 18}(e_s f \phi + e_u + e_d), \\
c_3={-1\over 72}(e_s + e_u + e_d), &  
c_4={-1\over 9}(e_s f \phi + (e_u+e_d)(f+1)/2),\\
c_5={1\over 18}(-2e_s f + 9e_s + (e_u+e_d)(9f+5)/2), & 
c_6={-1\over 18}(e_s f \phi + e_u + e_d)(7\kappa+\xi), \\
c_7={1\over 27}(-e_s f + 3e_s + (e_u+e_d)(5f-1)/2)\kappa_v, & 
c_8={-1\over 54}(e_s f \phi + (e_u+e_d)(f+1)/2)(7\kappa+\xi),\\
c_9={7\over 216}(e_s f \phi+ (e_u+e_d)(f+1)/2),&
c_{10}={-5\over 72}(e_s + (e_u+e_d)(f+1)/2),
\end{array}
\end{equation}
for ${\Xi^*}^0$:
\begin{equation}
\begin{array}{ll}
c_1={1\over 8}(2e_s + e_u), & 
c_2={-7\over 18}(2e_s f \phi + e_u), \\
c_3={-1\over 72}(2e_s + e_u), &  
c_4={-1\over 9} f (e_s f \phi + e_s \phi + e_u),\\
c_5={1\over 18}(5e_s f + 9e_s + 9e_u f - 2e_u), & 
c_6={-1\over 18}(2e_s f \phi + e_u)(7\kappa+\xi), \\
c_7={1\over 27} f (-e_s f + 5e_s + 3e_u f - e_u)\kappa_v, & 
c_8={-1\over 54} f (e_s f \phi + e_s \phi + e_u)(7\kappa+\xi),\\
c_9={7\over 216} f (e_s f \phi + e_s \phi + e_u),&
c_{10}={-5\over 72}(e_s f + e_u f + e_s),
\end{array}
\end{equation}
for $\Omega^-$:
\begin{equation}
\begin{array}{ll}
c_1={3\over 8}e_s, & 
c_2={-7\over 6}e_s f \phi, \\
c_3={-1\over 24}e_s, &  
c_4={-1\over 3}e_s f^2 \phi,\\
c_5={7\over 6}e_s f, & 
c_6={-1\over 6}e_s f \phi(7\kappa+\xi), \\
c_7={2\over 9}e_s f^2\kappa_v, & 
c_8={-1\over 17}e_s f^2 \phi(7\kappa+\xi),\\
c_9={7\over 72}e_s f^2 \phi,&
c_{10}={-5\over 24}e_s f.
\end{array}
\end{equation}

The sum rule from WE$_2$:
\begin{eqnarray}
& &
  c_1  L^{4/27} E_0 M^2
+ c_2  m_s \chi a L^{-12/27}
+ c_3  b L^{4/27} {1\over M^2} 
+ c_4  m_s a L^{4/27} {1\over M^2} 
+ c_5 m_s m^2_0 a L^{-10/27} {1\over M^4}
\nonumber \\ & &
={-1\over 9} \tilde{\lambda}^2_B
\left( {\mu_{\scriptscriptstyle B}\over M^2_B\,M^2} + A \right) e^{-M^2_B/M^2},
\label{we2}
\end{eqnarray}
where the coefficients for ${\Sigma^*}^+$ are:
\begin{equation}
\begin{array}{l}
c_1={-1\over 12}(e_s + 2e_u), 
c_2={-1\over 9}(e_s f \phi + 2e_u),  
c_3={-1\over 48}(e_s + 2e_u), 
c_4={ 1\over 3}(e_s + e_u f + e_u), \\
c_5={1\over 18}(e_s + e_u f + e_u),
\end{array}
\end{equation}
for ${\Xi^*}^0$:
\begin{equation}
\begin{array}{l}
c_1={-1\over 12}(2e_s + e_u), 
c_2={-1\over 9}(2e_s f \phi + e_u),  
c_3={-1\over 48}(2e_s + e_u), 
c_4={ 1\over 3}(e_s f + e_u f + e_s), \\
c_5={1\over 18}(e_s f + e_u f + e_s).
\end{array}
\end{equation}

The sum rule from WE$_3$:
\begin{eqnarray}
& &
  c_1  L^{4/27} E_1 M^4
+ c_2  m_s \chi a L^{-12/27} E_0 M^2
+ c_3  b L^{4/27} 
+ c_4 \chi a^2 L^{12/27} 
+ (c_5+c_6)  m_s a L^{4/27} 
\nonumber \\ & &
+ (c_7+c_8) a^2 L^{28/27} {1\over M^2}
+ c_9 \chi m^2_0 a^2 L^{-2/27} {1\over M^2}
+ c_{10} m_s  m^2_0 a L^{-10/27} {1\over M^2}
\nonumber \\ & &
={-7\over 18} \tilde{\lambda}^2_B
\left( {\mu_{\scriptscriptstyle B}\over M^2} + A \right) e^{-M^2_B/M^2},
\label{we3}
\end{eqnarray}
where the coefficients for ${\Sigma^*}^+$ are:
\begin{equation}
\begin{array}{ll}
c_1={-1\over 24}(e_s + 2e_u), & 
c_2={ 5\over 24}(e_s f \phi + 2e_u), \\
c_3={ 1\over 576}(e_s + 2e_u), &  
c_4={ 1\over 18}(e_s f \phi + e_u f_s + e_u),\\
c_5={1\over 36}(2e_s f - 3e_s - 3e_u f + e_u), & 
c_6={ 1\over 36}(e_s f \phi + 2e_u)(4\kappa+\xi), \\
c_7={1\over 54}( e_s f - 3e_s/2 - 2e_u f + e_u)\kappa_v & 
c_8={ 1\over 108}(e_s f \phi + e_u f + e_u)(4\kappa+\xi),\\
c_9={-7\over 432}(e_s f \phi + e_u f + e_u),&
c_{10}={ 1\over 48}(e_s + e_u f + e_u),
\end{array}
\end{equation}
for ${\Xi^*}^0$:
\begin{equation}
\begin{array}{ll}
c_1={-1\over 24}(2e_s + e_u), & 
c_2={ 5\over 24}(2e_s f \phi + e_u), \\
c_3={ 1\over 576}(2e_s + e_u), &  
c_4={ 1\over 18} f (e_s f \phi + e_s \phi + e_u),\\
c_5={1\over 36}(e_s f - 3e_s - 3e_u f + 2e_u), & 
c_6={ 1\over 36}(2e_s f \phi + e_u)(4\kappa+\xi), \\
c_7={1\over 54} f ( e_s f - 2e_s - 3e_u f/2 + e_u)\kappa_v & 
c_8={ 1\over 108} f (e_s f \phi + e_s \phi + e_u)(4\kappa+\xi),\\
c_9={-7\over 432} f (e_s f \phi + e_s \phi + e_u),&
c_{10}={ 1\over 48}(e_s f + e_u f + e_s).
\end{array}
\end{equation}

The sum rule from WE$_4$:
\begin{eqnarray}
& &
  c_1  L^{4/27} E_1 M^4
+ c_2  m_s \chi a L^{-12/27} E_0 M^2
+ c_3  b L^{4/27} 
+ (c_4+c_5)  m_s a L^{4/27} 
\nonumber \\ & &
+ (c_6+c_7) a^2 L^{28/27} {1\over M^2}
+ c_8  m^2_0 a^2 L^{14/27} {1\over M^4}
\nonumber \\ & &
={7\over 18} \tilde{\lambda}^2_B
\left( {\mu_{\scriptscriptstyle B}\over M^2} + A \right) e^{-M^2_B/M^2},
\label{we4}
\end{eqnarray}
where the coefficients for ${\Sigma^*}^+$ are:
\begin{equation}
\begin{array}{ll}
c_1={1\over 24}(e_s + 2e_u), & 
c_2={-5\over 36}(e_s f \phi + 2e_u), \\
c_3={-1\over 96}(e_s + 2e_u), &  
c_4={1\over 18}(e_s f + 6e_s + 6e_u f + 8e_u), \\
c_5={-1\over 72}(e_s f \phi + 2e_u)(10\kappa+\xi),& 
c_6={1\over 27}( e_s f + 3e_s/2 + 4e_u f + e_u)\kappa_v, \\
c_7={-1\over 108}(e_s f \phi + e_u f + e_u)(4\kappa+\xi), & 
c_8={-7\over 648}( e_s f + e_u f + e_u),
\end{array}
\end{equation}
for ${\Xi^*}^0$:
\begin{equation}
\begin{array}{ll}
c_1={1\over 24}(2e_s + e_u), & 
c_2={-5\over 36}(2e_s f \phi + e_u), \\
c_3={-1\over 96}(2e_s + e_u), &  
c_4={1\over 18}(8e_s f + 6e_s + 6e_u f + e_u), \\
c_5={-1\over 72}(2e_s f \phi + e_u)(10\kappa+\xi),& 
c_6={1\over 27} f ( e_s f + 4e_s + 3e_u f/2 + e_u)\kappa_v, \\
c_7={-1\over 108} f (e_s f \phi + e_s \phi + e_u)(4\kappa+\xi), & 
c_8={-7\over 648} f ( e_s f + e_s + e_u).
\end{array}
\end{equation}

The sum rule from WE$_5$:
\begin{eqnarray}
& &
  c_1  L^{4/27} E_1 M^4
+ c_2  m_s \chi a L^{-12/27} E_0 M^2
+ c_3  b L^{4/27} 
+ (c_4+c_5)  m_s a L^{4/27} 
+ c_6  \chi a^2 L^{12/27}
\nonumber \\ & &
+ (c_7+c_8) a^2 L^{28/27} {1\over M^2}
+ c_9 \chi m^2_0 a^2 L^{-2/27} {1\over M^2}
+ c_{10} m_s m^2_0 a L^{-10/27} {1\over M^2}
+ c_{11}  m^2_0 a^2 L^{14/27} {1\over M^4}
\nonumber \\ & &
={-7\over 18} \tilde{\lambda}^2_B
\left( {\mu_{\scriptscriptstyle B}\over M^2} + A \right) e^{-M^2_B/M^2},
\label{we5}
\end{eqnarray}
where the coefficients for ${\Sigma^*}^+$ are:
\begin{equation}
\begin{array}{ll}
c_1={-1\over 24}(e_s + 2e_u), & 
c_2={ 5\over 18}(e_s f \phi + 2e_u), \\
c_3={-1\over 144}(e_s + 2e_u), &  
c_4={ 1\over 6}(e_s f + e_s + e_u f + 3e_u), \\
c_5={ 1\over 24}(e_s f \phi + 2e_u)(2\kappa+\xi), & 
c_6={ 1\over 9}(e_s f \phi + e_u f + e_u), \\
c_7={2\over 27}( e_s f + e_u f + e_u)\kappa_v, & 
c_8={ 1\over 108}(e_s f \phi + e_u f + e_u)(4\kappa+\xi), \\
c_9={-7\over 216}( e_s f \phi + e_u f + e_u), &
c_{10}={ 1\over 24}(e_s +  e_u f + e_u), \\
c_{11}={-7\over 648}(e_s f +  e_u f + e_u),
\end{array}
\end{equation}
for ${\Xi^*}^0$:
\begin{equation}
\begin{array}{ll}
c_1={-1\over 24}(2e_s + e_u), & 
c_2={ 5\over 18}(2 e_s f \phi + e_u), \\
c_3={-1\over 144}(2e_s + e_u), &  
c_4={ 1\over 6}(3e_s f + e_s + e_u f + e_u), \\
c_5={ 1\over 24}(2e_s f \phi + e_u)(2\kappa+\xi), &
c_6={ 1\over 9} f (e_s f \phi + e_s \phi + e_u), \\
c_7={2\over 27} f ( e_s f + e_s + e_u)\kappa_v, & 
c_8={ 1\over 108} f (e_s f \phi + e_s\phi + e_u)(4\kappa+\xi), \\
c_9={-7\over 216} f ( e_s f \phi + e_s \phi + e_u), &
c_{10}={ 1\over 24}(e_s f +  e_u f + e_s), \\
c_{11}={-7\over 648} f (e_s f +  e_s + e_u).
\end{array}
\end{equation}

The sum rule from WE$_6$:
\begin{eqnarray}
& &
  c_1  m_s \chi a L^{-12/27} E_0 M^2
+ (c_2+c_3)  m_s a L^{4/27} 
+ (c_4+c_5) a^2 L^{28/27} {1\over M^2}
\nonumber \\ & &
+ c_6  m_s m^2_0 a L^{-10/27} {1\over M^2}
+ c_7  m^2_0 a^2 L^{14/27} {1\over M^4}
\nonumber \\ & &
={2\over 3} \tilde{\lambda}^2_B
\left( {\mu_{\scriptscriptstyle B}\over M^2} + A \right) e^{-M^2_B/M^2},
\label{we6}
\end{eqnarray}
where the coefficients for ${\Sigma^*}^+$ are:
\begin{equation}
\begin{array}{ll}
c_1={-2\over 3}(e_s f \phi + 2e_u),&
c_2={1\over 18}(e_s f + 2e_u), \\
c_3={ 1\over 72}(e_s f \phi + 2e_u)(2\kappa-\xi), & 
c_4={1\over 27} (e_s f + e_u f + e_u)\kappa_v,\\
c_5={ 1\over 108} (e_s f \phi + e_u f + e_u)(2\kappa-\xi),& 
c_6={1\over 36}(e_s + e_u f  + e_u),\\
c_7={-7\over 648}(e_s f + e_u f + e_u), 
\end{array}
\end{equation}
for ${\Xi^*}^0$:
\begin{equation}
\begin{array}{ll}
c_1={-2\over 3}(2 e_s f \phi + e_u),&
c_2={1\over 18}(2e_s f + e_u), \\
c_3={ 1\over 72}(2e_s f \phi + e_u)(2\kappa-\xi), & 
c_4={1\over 27} f (e_s f + e_s + e_u)\kappa_v,\\
c_5={ 1\over 108} f (e_s f \phi + e_s \phi + e_u)(2\kappa-\xi),& 
c_6={1\over 36}(e_s f + e_u f + e_s),\\
c_7={-7\over 648} f (e_s f + e_s + e_u). 
\end{array}
\end{equation}

The sum rule from WE$_7$ is identical to that from WE$_6$ after 
multiplying an overall sign on both sides.

The sum rule from WE$_8$:
\begin{eqnarray}
& &
  c_1  m_s \chi a L^{-12/27} E_0 M^2
+ c_2  b L^{4/27} 
+ c_3  \chi a^2 L^{12/27} 
+ (c_4+c_5)  m_s a L^{4/27} 
\nonumber \\ & &
+ (c_6+c_7) a^2 L^{28/27} {1\over M^2}
+ c_8  \chi m^2_0 a^2 L^{-2/27} {1\over M^2}
+ c_9  m_s m^2_0 a L^{-10/27} {1\over M^2}
+ c_{10}  m^2_0 a^2 L^{14/27} {1\over M^4}
\nonumber \\ & &
={-2\over 3} \tilde{\lambda}^2_B
\left( {\mu_{\scriptscriptstyle B}\over M^2} + A \right) e^{-M^2_B/M^2},
\label{we8}
\end{eqnarray}
where the coefficients for ${\Sigma^*}^+$ are:
\begin{equation}
\begin{array}{ll}
c_1={1\over 3}(e_s f \phi + 2e_u), & 
c_2={ 5\over 144}(e_s + 2e_u), \\
c_3={-2\over 9}(e_s f \phi + e_u f + e_u), &
c_4={-1\over 9}(5e_s f + 12e_s + 12e_u f + 22e_u),\\
c_5={1\over 36}(e_s f \phi + 2e_u)(14\kappa-\xi), & 
c_6={-1\over 27}(8e_s f + 6e_s + 20e_u f + 8e_u)\kappa_v,\\
c_7={ 1\over 54}(e_s f \phi + e_u f + e_u)(4\kappa+\xi),& 
c_8={ 7\over 108}(e_s f \phi + e_u f + e_u), \\
c_9={1\over 36}( e_s + e_u f + e_u),&
c_{10}={ 7\over 108}(e_s f + e_u f + e_u),
\end{array}
\end{equation}
for ${\Xi^*}^0$:
\begin{equation}
\begin{array}{ll}
c_1={1\over 3}(2e_s f \phi + e_u), & 
c_2={ 5\over 144}(2e_s + e_u), \\
c_3={-2\over 9} f (e_s f \phi + e_s \phi + e_u), &
c_4={-1\over 9}(22e_s f + 12e_s + 12e_u f + 5e_u),\\
c_5={1\over 36}(2e_s f \phi + e_u)(14\kappa-\xi), & 
c_6={-1\over 27} f (8e_s f + 20e_s + 6e_u f + 8e_u)\kappa_v,\\
c_7={ 1\over 54} f (e_s f \phi + e_s \phi + e_u)(4\kappa+\xi),& 
c_8={ 7\over 108} f (e_s f \phi + e_s\phi + e_u), \\
c_9={1\over 36}(e_s f + e_u f + e_s),&
c_{10}={ 7\over 108} f (e_s f + e_s + e_u).
\end{array}
\end{equation}

The sum rule from WE$_9$ has the same form as that from WE$_8$,
only with different $c_i$:
\begin{equation}
\begin{array}{ll}
c_1=(e_s f \phi + 2e_u), & 
c_2={-5\over 144}(e_s + 2e_u), \\
c_3={ 2\over 9}(e_s f \phi + e_u f + e_u), &
c_4={ 1\over 9}(4e_s f + 12e_s + 12e_u f + 20e_u),\\
c_5={1\over 18}(e_s f \phi + 2e_u)(-8\kappa+\xi), & 
c_6={1\over 9}(2e_s f + 2e_s + 6e_u f + 2e_u)\kappa_v,\\
c_7={-1\over 9}(e_s f \phi + e_u f + e_u)\kappa,& 
c_8={-7\over 108}(e_s f \phi + e_u f + e_u), \\
c_9={-1\over 12}(e_s + e_u f + e_u),&
c_{10}={-7\over 162}(e_s f + e_u f + e_u),
\end{array}
\end{equation}
for ${\Xi^*}^0$:
\begin{equation}
\begin{array}{ll}
c_1=(2e_s f \phi + e_u), & 
c_2={-5\over 144}(2e_s + e_u), \\
c_3={ 2\over 9} f (e_s f \phi + e_s \phi + e_u), &
c_4={ 1\over 9}(20e_s f + 12e_s + 12e_u f + 4e_u),\\
c_5={1\over 18}(2 e_s f \phi + e_u)(-8\kappa+\xi), & 
c_6={1\over 9} f (2e_s f + 6e_s + 2e_u f + 2e_u)\kappa_v,\\
c_7={-1\over 9} f (e_s f \phi + e_s \phi + e_u)\kappa,& 
c_8={-7\over 108} f (e_s f \phi + e_s \phi + e_u), \\
c_9={-1\over 12}(e_s f + e_u f + e_s),&
c_{10}={-7\over 162} f (e_s f + e_s + e_u).
\end{array}
\end{equation}

The sum rule from WO$_1$:
\begin{eqnarray}
& &
  c_1 \chi a E_1 M^4
+ c_2 m_s L^{16/27} E_1 M^4
+ (c_3+c_4) a L^{16/27} E_0 M^2
+ c_5 m^2_0 a L^{2/27}
\nonumber \\ & &
+ c_6 \chi a b
+ c_7 m_s \chi a^2
+ c_8 a b L^{16/27} {1\over M^2}
+ (c_9+c_{10}) m_s a^2 L^{16/27} {1\over M^2}
\nonumber \\ & &
={1\over 2} \tilde{\lambda}^2_B
\left( {\mu_{\scriptscriptstyle B}\,M_B\over M^2} + A \right) e^{-M^2_B/M^2},
\label{wo1}
\end{eqnarray}
where the coefficients for ${\Sigma^*}^+$ are:
\begin{equation}
\begin{array}{ll}
c_1={-1\over 6}(e_s f \phi + 2e_u), & 
c_2={ 1\over 6}(e_s + 2e_u), \\
c_3={-5\over 27}(e_s f + 2e_u), &  
c_4={ 5\over 432}(e_s f \phi + 2e_u)(-8\kappa+7\xi),\\
c_5={1\over 12}(e_s + e_u f + e_u), & 
c_6={ 1\over 96}(e_s f \phi + 2e_u), \\
c_7={-2\over 9}(e_s f \phi + e_u f + e_u), & 
c_8={-1\over 216}(e_s + e_u f + e_u),\\
c_9={1\over 27}(-2e_s f + 3e_s + 4e_u f - 2e_u)\kappa_v,&
c_{10}={-5\over 432}(e_s f \phi + e_u f + e_u)(8\kappa+11\xi),
\end{array}
\end{equation}
for ${\Xi^*}^0$:
\begin{equation}
\begin{array}{ll}
c_1={-1\over 6}(2e_s f \phi + e_u), & 
c_2={ 1\over 6}(2 e_s + e_u), \\
c_3={-5\over 27}(2e_s f + e_u), &  
c_4={ 5\over 432}(2e_s f \phi + e_u)(-8\kappa+7\xi),\\
c_5={1\over 12}(e_s f + e_u f + e_s), & 
c_6={ 1\over 96}(2e_s f \phi + e_u), \\
c_7={-2\over 9} f (e_s f \phi + e_s \phi + e_u), & 
c_8={-1\over 216}(e_s f + e_u f + e_s),\\
c_9={1\over 27} f (-2e_s f + 4e_s + 3e_u f - 2e_u)\kappa_v,&
c_{10}={-5\over 432} f (e_s f \phi + e_s \phi + e_u)(8\kappa+11\xi).
\end{array}
\end{equation}

The sum rule from WO$_2$:
\begin{eqnarray}
& &
  c_1 \chi a E_0 M^2
+ (c_2+c_3)  a L^{16/27}
+ c_4 m_s \chi a^2
+ c_5 \chi a b {1\over M^2}
+ (c_6+c_7) m_s a^2 L^{16/27} {1\over M^4}
\nonumber \\ & &
={-1\over 9} \tilde{\lambda}^2_B
\left( {\mu_{\scriptscriptstyle B}\over M_B\,M^2} + A \right) e^{-M^2_B/M^2},
\label{wo2}
\end{eqnarray}
where the coefficients for ${\Sigma^*}^+$ are:
\begin{equation}
\begin{array}{ll}
c_1={ 1\over 9}(e_s f \phi + 2e_u), & 
c_2={ 5\over 27}(e_s f + 2e_u), \\
c_3={ 1\over 54}(e_s f \phi + 2e_u)(5\kappa+2\xi), &  
c_4={ -2\over 9}(e_s f \phi + e_u f + e_u),\\
c_5={1\over 216}(e_s f \phi + 2e_u), & 
c_6={-2\over 9}(e_s f + e_u f + e_u)\kappa_v, \\
c_7={-1\over 9}(e_s f \phi + e_u f + e_u)\kappa, 
\end{array}
\end{equation}
for ${\Xi^*}^0$:
\begin{equation}
\begin{array}{ll}
c_1={ 1\over 9}(2e_s f \phi + e_u), & 
c_2={ 5\over 27}(2e_s f + e_u), \\
c_3={ 1\over 54}(2e_s f \phi + e_u)(5\kappa+2\xi), &  
c_4={ -2\over 9} f (e_s f \phi + e_s \phi + e_u),\\
c_5={1\over 216}(2e_s f \phi + e_u), & 
c_6={-2\over 9} f (e_s f + e_s + e_u)\kappa_v, \\
c_7={-1\over 9} f (e_s f \phi + e_s\phi + e_u)\kappa. 
\end{array}
\end{equation}

The sum rule from WO$_3$:
\begin{eqnarray}
& &
  c_1 \chi a E_1 M^4
+ (c_2+c_3)  a L^{16/27} E_0 M^2
+ c_4 m^2_0 a L^{2/27}
+ c_5 \chi a b
+ c_6 m_s\chi a^2
\nonumber \\ & &
+ c_7  a b L^{16/27} {1\over M^2}
+ (c_8+c_9) m_s a^2 L^{16/27} {1\over M^2}
\nonumber \\ & &
={-7\over 18} \tilde{\lambda}^2_B
\left( {\mu_{\scriptscriptstyle B}\,M_B\over M^2} + A \right) e^{-M^2_B/M^2},
\label{wo3}
\end{eqnarray}
where the coefficients for ${\Sigma^*}^+$ are:
\begin{equation}
\begin{array}{ll}
c_1={ 5\over 72}(e_s f \phi + 2e_u), &
c_2={1\over 72}(7e_s f + 6 e_s + 6 e_u f + 20e_u), \\
c_3={ 1\over 288}(e_s f \phi + 2e_u)(2\kappa-11\xi), &  
c_4={-1\over 12}(e_s + e_u f + e_u),\\
c_5={-11\over 1728}(e_s f \phi + 2e_u), & 
c_6={ 1\over 12}(e_s f \phi + e_u f + e_u), \\
c_7={5\over 1728}(e_s + e_u f + e_u),& 
c_8={-1\over 36}(e_s f + 3e_s + 7e_u f + e_u)\kappa_v,\\
c_9={ 1\over 432}(e_s f \phi + e_u f + e_u)(12\kappa+7\xi),
\end{array}
\end{equation}
for ${\Xi^*}^0$:
\begin{equation}
\begin{array}{ll}
c_1={ 5\over 72}(2e_s f \phi + e_u), &
c_2={1\over 72}(20e_s f + 6 e_s + 6 e_u f + 7e_u), \\
c_3={ 1\over 288}(2e_s f \phi + e_u)(2\kappa-11\xi), &  
c_4={-1\over 12}(e_s f + e_u f + e_s),\\
c_5={-11\over 1728}(2e_s f \phi + e_u), & 
c_6={ 1\over 12} f (e_s f \phi + e_s\phi + e_u), \\
c_7={5\over 1728}(e_s f + e_u f + e_s),& 
c_8={-1\over 36} f (e_s f + 7e_s + 3e_u f + e_u)\kappa_v,\\
c_9={ 1\over 432} f (e_s f \phi + e_s \phi + e_u)(12\kappa+7\xi).
\end{array}
\end{equation}

The sum rule from WO$_4$:
\begin{eqnarray}
& &
  c_1  m_s L^{-8/27} E_0 M^2
+ (c_2+c_3)  a L^{16/27} 
+ c_4 m_s\chi a^2
+ c_5  m^2_0 a L^{2/27} {1\over M^2}
+ c_6  \chi a b {1\over M^2}
\nonumber \\ & &
+ c_7  a b L^{16/27} {1\over M^4}
+ (c_8+c_9)  m_s a^2 L^{16/27} {1\over M^4}
\nonumber \\ & &
={7\over 18} \tilde{\lambda}^2_B
\left( {\mu_{\scriptscriptstyle B}\over M_B\,M^2} + A \right) e^{-M^2_B/M^2},
\label{wo4}
\end{eqnarray}
where the coefficients for ${\Sigma^*}^+$ are:
\begin{equation}
\begin{array}{ll}
c_1={1\over 6}(e_s + 2e_u), & 
c_2={ 1\over 6}(e_s + e_u f + e_u), \\
c_3={-1\over 144}(e_s f \phi + 2e_u)(12\kappa+\xi), &  
c_4={ 1\over 9}(e_s f \phi + e_u f + e_u),\\
c_5={-1\over 12}(e_s + e_u f + e_u), & 
c_6={-1\over 288}(e_s f \phi + 2e_u), \\
c_7={1\over 432}(e_s + e_u f + e_u),& 
c_8={-1\over 18}(e_s + 2e_u f)\kappa_v,\\
c_9={1\over 432}(e_s f \phi + e_u f + e_u)(12\kappa+\xi),
\end{array}
\end{equation}
for ${\Xi^*}^0$:
\begin{equation}
\begin{array}{ll}
c_1={1\over 6}(2e_s + e_u), & 
c_2={ 1\over 6}(e_s f + e_u f + e_s), \\
c_3={-1\over 144}(2e_s f \phi + e_u)(12\kappa+\xi), &  
c_4={ 1\over 9} f (e_s f \phi + e_s \phi + e_u),\\
c_5={-1\over 12}(e_s f + e_u f + e_s), & 
c_6={-1\over 288}(2e_s f \phi + e_u), \\
c_7={1\over 432}(e_s f + e_u f + e_s),& 
c_8={-1\over 18} f (2e_s + e_u f)\kappa_v,\\
c_9={1\over 432} f (e_s f \phi + e_s \phi + e_u)(12\kappa+\xi).
\end{array}
\end{equation}

The sum rule from WO$_5$ is identical to that from WO$_4$ 
after multiplying an overall sign on both sides.

The sum rule from WO$_6$:
\begin{eqnarray}
& &
  c_1  \chi a E_1 M^4
+ c_2  m_s L^{-8/27} E_1 M^4
+ (c_3+c_4)  a L^{16/27} E_0 M^2
\nonumber \\ & &
+ c_5  m^2_0 a L^{2/27}
+ c_6 \chi a b
+ c_7  a b L^{16/27} {1\over M^2}
+ (c_8+c_9)  m_s a^2 L^{16/27} {1\over M^2}
\nonumber \\ & &
={2\over 3} \tilde{\lambda}^2_B
\left( {\mu_{\scriptscriptstyle B}\,M_B\over M^2} + A \right) e^{-M^2_B/M^2},
\label{wo6}
\end{eqnarray}
where the coefficients for ${\Sigma^*}^+$ are:
\begin{equation}
\begin{array}{ll}
c_1={-1\over 9}(e_s f \phi + 2e_u), & 
c_2={-1\over 3}(e_s + 2e_u), \\
c_3={-1\over 54}(11e_s f + 18e_s + 18e_u f + 40e_u), & 
c_4={ 1\over 216}(e_s f \phi + 2e_u)(14\kappa+23\xi), \\
c_5={ 1\over 6}(e_s + e_u f + e_u),& 
c_6={5\over 432}(e_s f \phi + 2e_u), \\
c_7={-1\over 432}(e_s + e_u f + e_u), & 
c_8={1\over 27}(5e_s f + 3e_s + 11e_u f + 5e_u)\kappa_v,\\
c_9={ 1\over 54}(e_s f \phi + e_u f + e_u)(2\kappa-3\xi),
\end{array}
\end{equation}
for ${\Xi^*}^0$:
\begin{equation}
\begin{array}{ll}
c_1={-1\over 9}(2e_s f \phi + e_u), & 
c_2={-1\over 3}(2e_s + e_u), \\
c_3={-1\over 54}(40e_s f + 18e_s + 18e_u f + 11e_u), & 
c_4={ 1\over 216}(2e_s f \phi + e_u)(14\kappa+23\xi), \\
c_5={ 1\over 6}(e_s f + e_u f + e_s),& 
c_6={5\over 432}(2 e_s f \phi + e_u), \\
c_7={-1\over 432}(e_s f + e_u f + e_s), & 
c_8={1\over 27} f (5e_s f + 11e_s + 3e_u f + 5e_u)\kappa_v,\\
c_9={ 1\over 54} f (e_s f \phi + e_s\phi + e_u)(2\kappa-3\xi).
\end{array}
\end{equation}

The sum rule from WO$_7$ has the same form as that from WO$_6$
after multiplying an overall sign on both sides.
They only differ in $c_4$ and $c_9$ for ${\Sigma^*}^+$:
\begin{equation}
c_4={ 1\over 216}(e_s f \phi + 2e_u)(14\kappa+29\xi), 
\hspace{2mm}
c_9={ 1\over 108}(e_s f \phi + e_u f + e_u)(-4\kappa+5\xi),
\end{equation}
for ${\Xi^*}^0$:
\begin{equation}
c_4={ 1\over 216}(2e_s f \phi + e_u)(14\kappa+29\xi), 
\hspace{2mm}
c_9={ 1\over 108} f (e_s f \phi + e_s\phi + e_u)(-4\kappa+5\xi).
\end{equation}

The sum rule from WO$_8$:
\begin{eqnarray}
& &
  c_1  \chi a E_0 M^2
+ c_2  m_s L^{-8/27} E_0 M^2
+ (c_3+c_4)  a L^{16/27} 
+ c_5  m^2_0 a L^{2/27} {1\over M^2}
\nonumber \\ & &
+ c_6 \chi a b {1\over M^2}
+ c_7  a b L^{16/27} {1\over M^4}
+ (c_8+c_9)  m_s a^2 L^{16/27} {1\over M^4}
\nonumber \\ & &
={-2\over 3} \tilde{\lambda}^2_B
\left( {\mu_{\scriptscriptstyle B}\over M_B\,M^2} + A \right) e^{-M^2_B/M^2},
\label{wo8}
\end{eqnarray}
where the coefficients for ${\Sigma^*}^+$ are:
\begin{equation}
\begin{array}{ll}
c_1={-1\over 9}(e_s f \phi + 2e_u), & 
c_2={-1\over 3}(e_s + 2e_u), \\
c_3={-1\over 27}(5e_s f + 9e_s + 9e_u f + 19e_u), & 
c_4={ 1\over 108}(e_s f \phi + 2e_u)(8\kappa-\xi), \\
c_5={ 1\over 6}(e_s + e_u f + e_u),& 
c_6={1\over 432}(e_s f \phi + 2e_u), \\
c_7={-1\over 216}(e_s + e_u f + e_u), & 
c_8={1\over 9}(2e_s f + e_s + 4e_u f + 2e_u)\kappa_v, \\
c_9={ 1\over 108}(e_s f \phi + e_u f + e_u)(6\kappa-\xi),
\end{array}
\end{equation}
for ${\Xi^*}^0$:
\begin{equation}
\begin{array}{ll}
c_1={-1\over 9}(2e_s f \phi + e_u), & 
c_2={-1\over 3}(2e_s + e_u), \\
c_3={-1\over 27}(19e_s f + 9e_s + 9e_u f + 5e_u), & 
c_4={ 1\over 108}(2e_s f \phi + e_u)(8\kappa-\xi), \\
c_5={ 1\over 6}(e_s f + e_u f + e_s),& 
c_6={1\over 432}(2e_s f \phi + e_u), \\
c_7={-1\over 216}(e_s f + e_u f + e_s), & 
c_8={1\over 9} f (2e_s f + 4e_s + e_u f + 2e_u)\kappa_v, \\
c_9={ 1\over 108} f (e_s f \phi + e_s \phi + e_u)(6\kappa-\xi).
\end{array}
\end{equation}

The sum rule from WO$_9$ has the same form as that from WO$_8$.
They only differ in $c_4$ and $c_9$ for ${\Sigma^*}^+$:
\begin{equation}
c_4={ 1\over 27}(e_s f \phi + 2e_u)(2\kappa-\xi), 
\hspace{2mm}
c_9={ 1\over 18}(e_s f \phi + e_u f + e_u)\kappa,
\end{equation}
for ${\Xi^*}^0$:
\begin{equation}
c_4={ 1\over 27}(2e_s f \phi + e_u)(2\kappa-\xi), 
\hspace{2mm}
c_9={ 1\over 18} f (e_s f \phi + e_s\phi + e_u)\kappa.
\end{equation}
%



%
\begin{table}[p]
\caption{Monte-Carlo analysis of the QCD sum rules 
for the magnetic moment of $\Delta^{++}$ and $\Omega^{-}$.
The six columns correspond to, from left to right: 
the sum rule that has a valid Borel region, 
the Borel region determined by the 10\%-50\% criteria, 
the percentage contribution of the excited states and transitions 
to the phenomenological side at the lower end of the Borel region 
(it increases to 50\% at the upper end),  
the continuum threshold,
the transition strength,
the magnetic moment in nuclear magnetons.
The uncertainties in each sum rule were obtained from 
consideration of 1000 QCD parameter sets.}
\label{tabdo}
\begin{tabular}{rccccc}
Sum Rule & Region & Cont      & $w$ & A & $\mu_{\scriptscriptstyle B}$ \\
& (GeV) & (\%) & (GeV) & (GeV$^{-2}$)   &  $(\mu_{\scriptscriptstyle N})$ 
\\ \hline
$\Delta^{++}$: \hspace{1mm} 
WE$_1$
& 0.70 to 1.53 & 1.7& 1.65      &-0.28 $\pm$ 0.52 & 7.76 $\pm$ 2.67 \\
WE$_3$
& 1.04 to 1.42 & 19 & 1.65      &0.20 $\pm$ 0.20 & 3.06 $\pm$ 1.14\\
WE$_4$
& 0.675 to 1.56 &  5 & 1.65      &-0.35$\pm$ 0.37 & 3.34 $\pm$ 1.44 \\
WE$_5$
& 0.765 to 1.47 & 8.5& 1.65      & 0.53$\pm$ 0.81 & 3.56$\pm$ 3.49 \\
\hline
$\Omega^{-}$: \hspace{1mm} 
WE$_1$
&0.592 to 1.70 & 2.3& 2.30      &-0.12$\pm$ 0.11 & -2.66 $\pm$ 0.88 \\
WE$_2$
&0.872 to 1.53 & 20 & 2.30      &-0.26$\pm$ 0.20 & -5.31 $\pm$ 3.66\\
WE$_3$
&0.885 to 1.68 & 8.5& 2.30      &-0.09$\pm$ 0.04 & -1.24 $\pm$ 0.51\\
WE$_4$
&0.60  to 1.72 &  2 & 2.30      &-0.03$\pm$ 0.05 &-1.24 $\pm$ 0.24 \\
WE$_5$
&0.747 to 1.66 & 7.4& 2.30      &-0.14$\pm$ 0.14 &-1.32 $\pm$ 1.08 \\
WE$_6$
&0.59  to 2.32 & 0.86& 2.30      &-0.01 $\pm$ 0.02  & -1.14 $\pm$ 0.40 \\
WE$_8$
&0.69  to 2.60 & 3.3& 2.30      & 0.03 $\pm$ 0.03 & -0.65 $\pm$ 1.22 \\
WO$_1$
&0.663 to 1.26  &12 & 2.30      &-0.32$\pm$ 0.42 &-0.65 $\pm$ 1.22 \\
WO$_2$
& 1.06 to 1.43  & 31& 2.30      &-0.62$\pm$ 0.18 & -4.94 $\pm$ 5.58\\
WO$_4$
&0.836 to 2.22  &7.4& 2.30      &-0.03$\pm$ 0.01 & -0.70 $\pm$  0.24    
\end{tabular}
\end{table}
%
%
\begin{table}[p]
\caption{Same as Table~\protect\ref{tabdo}, but for 
${\Sigma^*}^{+}$, ${\Sigma^*}^{0}$ and ${\Sigma^*}^{-}$.
The presence of a second row in a specific sum rule 
indicates that the continuum threshold was successfully searched.}
\label{tabsig}
\begin{tabular}{rccccc}
Sum Rule & Region & Cont      & $w$ & A & $\mu_{\scriptscriptstyle B}$ \\
& (GeV) & (\%) & (GeV) & (GeV$^{-2}$) &  $(\mu_{\scriptscriptstyle N})$         
\\ \hline
${\Sigma^*}^{+}$: \hspace{1mm} 
WE$_1$
&0.853 to 1.445& 11 & 1.80      & 0.28$\pm$ 0.18 & 2.96 $\pm$ 1.41 \\
WE$_3$
&0.996 to 1.39 & 23 & 1.80      & 0.23$\pm$ 0.08 & 1.49 $\pm$ 0.79  \\
WE$_4$
&0.622 to 1.61 &  1 & 1.80      &-0.05$\pm$ 0.10 & 1.74 $\pm$ 0.42 \\
WE$_5$
&0.715 to 1.45 & 10 & 1.80      & 0.34$\pm$ 0.35 & 1.82 $\pm$ 1.94 \\
&0.715 to 1.45 & 6  & 2.65 $\pm$ 5.96 & 0.26$\pm$ 0.49 & 1.71 $\pm$ 1.96 \\
WE$_6$
&0.575 to 1.96 & 0.2& 1.80      &-0.01 $\pm$ 0.08 & 2.10 $\pm$  0.79 \\
&0.575 to 1.96 & 0.9& 1.56$\pm$ 0.11 &-0.06 $\pm$ 0.04 & 2.00 $\pm$ 0.68 \\
WE$_8$
&0.79  to 2.36 & 13 & 1.80      &-0.17 $\pm$ 0.08 & 1.09 $\pm$ 0.71 \\
&0.79  to 2.36 & 15 & 1.52$\pm$ 5.39&-0.21$\pm$ 0.08 & 1.08 $\pm$ 0.67 \\
WO$_4$
&0.89  to 1.46 & 23 & 1.80      & 0.07 $\pm$ 0.06 & 0.39 $\pm$ 0.48 \\
\hline
${\Sigma^*}^{0}$: \hspace{1mm} 
WE$_5$
&0.577 to 1.95 & 2.8& 1.80      &0.01 $\pm$ 0.01 & 0.19 $\pm$ 0.13 \\
WE$_8$
&0.639 to 1.70 & 9.4& 1.80      &0.03 $\pm$ 0.01  & -0.18  $\pm$ 0.06 \\
WO$_2$
&0.846 to 1.38 & 18 & 1.80      &0.11 $\pm$ 0.09 & 1.00 $\pm$ 0.96\\
WO$_8$
&0.662 to 1.66 &  5 & 1.80      &-0.01$\pm$ 0.01 & -0.30 $\pm$ 0.18\\
WO$_9$
&0.627 to 1.73 & 3.4& 1.80      &-0.01  $\pm$ 0.01  & -0.33 $\pm$ 0.19 \\
\hline
${\Sigma^*}^{-}$: \hspace{1mm} 
WE$_1$
&0.662 to 1.54 & 1  & 1.80      &-0.05$\pm$ 0.19 & -3.34 $\pm$ 1.33 \\
WE$_3$
&0.926 to 1.42 & 16 & 1.80      &-0.17$\pm$ 0.08 & -1.42 $\pm$ 0.71  \\
WE$_4$
&0.602 to 1.61 & 1.3& 1.80      &0.06 $\pm$ 0.10 & -1.70 $\pm$ 0.38  \\
WE$_5$
&0.735 to 1.37 & 13 & 1.80      &-0.33$\pm$ 0.36 & -1.40 $\pm$ 1.74  \\
WE$_6$
&0.588 to 1.97 & 0.2& 1.80      &0.01 $\pm$ 0.07 & -1.72  $\pm$ 0.63 \\
WE$_8$
&0.71  to 2.51 &9.4 & 1.80      &0.15 $\pm$ 0.07 & -1.22 $\pm$  0.65 \\
WO$_1$
&0.618 to 1.05 &10  & 1.80      &-0.34 $\pm$ 0.77 & -0.66 $\pm$ 1.45 \\
WO$_4$
&0.89  to 1.57 & 19 & 1.80      &-0.08 $\pm$ 0.05 & -0.54 $\pm$ 0.39 \\
\end{tabular}
\end{table}
%
%
\begin{table}[p]
\caption{Same as Table~\protect\ref{tabdo}, but for
${\Xi^*}^0$ and ${\Xi^*}^-$.}
\label{tabxi}
\begin{tabular}{rccccc}
Sum Rule & Region & Cont      & $w$ & A & $\mu_{\scriptscriptstyle B}$ \\
& (GeV) & (\%) & (GeV) & (GeV$^{-x}$)  &  $(\mu_{\scriptscriptstyle N})$         
\\ \hline
${\Xi^*}^0$: \hspace{1mm} 
WE$_8$
&0.636 to 1.55 & 13 & 2.00      & 0.08 $\pm$ 0.03 & -0.35 $\pm$ 0.12 \\
WO$_2$
&0.977 to 1.25  &13 & 2.00      & 0.11$\pm$ 0.13 & 2.25  $\pm$ 1.92 \\
WO$_8$
&0.654 to 1.85  &3.7& 2.00      &-0.02$\pm$ 0.02 & -0.62 $\pm$ 0.34\\
WO$_9$
&0.621 to 1.91  &2.8& 2.00      &-0.02$\pm$ 0.02 & -0.69 $\pm$ 0.35 \\
\hline
${\Xi^*}^-$: \hspace{1mm} 
WE$_1$
&0.628 to 1.61 &1.3 & 2.00      &-0.07$\pm$ 0.14 & -2.88 $\pm$ 1.02  \\
WE$_2$
&0.898 to 1.12 & 37 & 2.00      &-0.30$\pm$ 0.83 & -3.65 $\pm$ 5.68  \\
WE$_3$
&0.906 to 1.53 & 12 & 2.00      &-0.12$\pm$ 0.05 & -1.25  $\pm$ 0.55 \\
WE$_4$
&0.6   to 1.66 & 0.3& 2.00      &0.0004$\pm$ 0.07 & -1.38 $\pm$ 0.27  \\
WE$_5$
&0.74  to 1.50 & 10 & 2.00      &-0.22$\pm$ 0.21 & -1.27 $\pm$ 1.30  \\
WE$_6$
&0.59  to 2.11 &0.5 & 2.00      &-0.006 $\pm$ 0.04 & -1.38  $\pm$ 0.48  \\
WE$_8$
&0.70  to 2.54 & 6.3& 2.00      &0.07  $\pm$ 0.05 & -0.88 $\pm$ 0.47 \\
WO$_1$
&0.641 to 1.12  & 13& 2.00      &-0.35$\pm$ 0.57 & -0.58 $\pm$ 1.26 \\
WO$_4$
&0.863 to 1.86  &12 & 2.00      &-0.05$\pm$ 0.02 & -0.60 $\pm$ 0.29 
\end{tabular}
\end{table}
%
%
\begin{table}[p]
\caption{Comparisons of decuplet baryon magnetic moments from
various calculations: this work (QCDSR),
lattice QCD (Latt)~\protect\cite{Derek92},
chiral perturbation theory ($\chi$PT)~\protect\cite{Butler94},
light-cone relativistic quark model
(RQM)~\protect\cite{Schlumpf93},
non-relativistic quark model (NQM)~\protect\cite{PDG92},
chiral quark-soliton model ($\chi$QSM)~\protect\cite{Kim97}.
All results are in units of nuclear magnetons.}
\label{comp}
\begin{tabular}{lccccccc}
Baryon & Exp.  & QCDSR & Latt & $\chi$PT &  RQM & NQM & $\chi$QSM
\\ \hline
$\Delta^{++}$
&4.5 $\pm$ 1.0  &  4.13 $\pm$ 1.30 & 4.91 $\pm$ 0.61 
& 4.0 $\pm$ 0.4 & 4.76 & 5.56 & 4.73 \\
$\Delta^{+}$
&  &  2.07 $\pm$ 0.65    & 2.46 $\pm$ 0.31 
& 2.1   $\pm$ 0.2  & 2.38 & 2.73 & 2.19 \\
$\Delta^{0}$
& $\approx$ 0 &  0.00   & 0.00 & -0.17 $\pm$ 0.04 & 0.00 &-0.09 & -0.35\\
$\Delta^{-}$
&   & -2.07 $\pm$ 0.65     & -2.46 $\pm$ 0.31
& -2.25 $\pm$ 0.25 &-2.38 &-2.92 & -2.90\\
${\Sigma^*}^{+}$
&   &  2.13 $\pm$ 0.82          & 2.55 $\pm$ 0.26 
& 2.0   $\pm$ 0.2  & 1.82 & 3.09 & 2.52 \\
${\Sigma^*}^{0}$
&   & -0.32 $\pm$ 0.15          & 0.27 $\pm$ 0.05 
& -0.07 $\pm$ 0.02 &-0.27 & 0.27 & -0.08\\
${\Sigma^*}^{-}$
&   & -1.66 $\pm$ 0.73          &-2.02 $\pm$ 0.18 
& -2.2  $\pm$ 0.2  &-2.36 &-2.56 & -2.69\\
${\Xi^*}^0$
&   & -0.69 $\pm$ 0.29         & 0.46 $\pm$ 0.07 
&  0.1  $\pm$ 0.04 &-0.60 & 0.63 & 0.19 \\
${\Xi^*}^-$
&   &  -1.51$\pm$ 0.52          &-1.68 $\pm$ 0.12 
& -2.0  $\pm$ 0.2  &-2.41 & -2.2 & -2.48\\
$\Omega^{-}$
& -2.024 $\pm$ 0.056 & -1.49 $\pm$ 0.45 & -1.40 $\pm$ 0.10 
& \mbox{input}           &-2.48 & -1.84 & -2.27\\
\end{tabular}
\end{table}
%


%
%
%
%
%
%
%
%
%
%
%
%
%
%
%

%
%
%
%

\end{document}